\documentclass[trackchanges, twocolumn]{aastex701}
\usepackage{booktabs}
\usepackage{xcolor}
\usepackage{upgreek}
\usepackage[upgreek]{mathastext}
\definecolor{BLUE}{named}{blue}

\begin{document}

\title{Dwarfs of Fire and Ice: UV-Irradiation-Driven Thermal Inversion, Disequilibrium Chemistry, and a Cloud-Free Atmosphere in a Strongly Irradiated Brown Dwarf Probed by JWST/NIRSpec Phase-Resolved Spectroscopy}

\author[orcid=0009-0003-6759-7067,sname='Shin']{Lael Shin}
\affiliation{Steward Observatory and Department of Astronomy, University of Arizona, 933 N. Cherry Ave., Tucson, AZ 85721, USA}
\email[show]{lshin@arizona.edu}  

\author[orcid=0000-0003-3714-5855,sname='Apai']{Daniel Apai} 
\affiliation{Steward Observatory and Department of Astronomy, University of Arizona, 933 N. Cherry Ave., Tucson, AZ 85721, USA}
\affiliation{Lunar and Planetary Laboratory, University of Arizona, 933 N. Cherry Ave., Tucson, AZ 85721, USA}
\affiliation{James C. Wyant College of Optical Sciences, University of Arizona, Tucson, AZ 85721, USA}
\email[show]{apai@arizona.edu}

\author[orcid=0000-0003-2969-6040, sname='Zhou']{Yifan Zhou}
\affiliation{Department of Astronomy, University of Virginia, Charlottesville, VA 22904, USA}
\email{yzhou@virginia.edu}

\author[orcid=0000-0003-3667-8633, gname={Joshua}, sname={Lothringer}]{Joshua D. Lothringer}
\affiliation{Space Telescope Science Institute, 3700 San Martin Drive, Baltimore, MD 21218, USA}
\email{jlothrin@stsci.edu}

\author[orcid=0009-0005-9531-7733, sname='Broski-Laing']{Daphne Broski-Laing}
\affiliation{Department of Astronomy, University of Virginia, Charlottesville, VA 22904, USA}
\email{daphne.broski.laing@virginia.edu}

\author[orcid=0000-0003-2478-0120, gname={Sarah}, sname={Casewell}]{Sarah L. Casewell}
\affiliation{Centre for Exoplanet Research, School of Physics and Astronomy, University of Leicester, Leicester LE1 7RH, UK}
\email{slc25@leicester.ac.uk}

\author[orcid=0000-0002-8808-4282, gname={Siyi}, sname={Xu}]{Siyi Xu}
\affiliation{Gemini Observatory / NSF NOIRLab, 950 N.\ Cherry Avenue, Tucson, AZ 85719, USA}
\email{siyi.xu@noirlab.edu}

\author[orcid=0000-0003-1487-6452, gname={Ben}, sname={Lew}]{Ben W. P. Lew}
\affiliation{Bay Area Environmental Research Institute, Moffett Field, CA 94035, USA}
\affiliation{NASA Ames Research Center, Moffett Field, CA 94035, USA}
\email{weipenglew@arizona.edu}

\author[orcid=0000-0003-2278-6932, gname={Xianyu}, sname={Tan}]{Xianyu Tan}
\affiliation{Tsung-Dao Lee Institute \& School of Physics and Astronomy, Shanghai Jiao Tong University, Shanghai 201210, China}
\email{xianyut@sjtu.edu.cn}

\author[orcid=0000-0003-0825-4876, gname={Jenni}, sname={French}]{Jenni R. French}
\affiliation{School of Physics and Astronomy, University of Birmingham, Birmingham B15 2TT, UK}
\email{j.french.4@bham.ac.uk}

\author[orcid=0000-0001-9521-6258, gname={Vivien}, sname={Parmentier}]{Vivien Parmentier}
\affiliation{Laboratoire Lagrange, Observatoire de la C\^{o}te d'Azur, Universit\'{e} C\^{o}te d'Azur, Nice, France}
\email{vivien.parmentier@oca.eu}

\author[orcid=0000-0002-5251-2943, gname={Mark}, sname={Marley}]{Mark S. Marley}
\affiliation{Lunar and Planetary Laboratory, University of Arizona, Tucson, AZ 85721, USA}
\email{marley@lpl.arizona.edu}

\author[orcid=0000-0002-4321-4581, gname={Laura}, sname={Mayorga}]{L. C. Mayorga}
\affiliation{Johns Hopkins University Applied Physics Laboratory, Laurel, MD 20723, USA}
\email{Laura.Mayorga@jhuapl.edu}

\begin{abstract}
We present time-resolved JWST/NIRSpec PRISM spectrophotometry of the white dwarf (WD)--brown dwarf (BD) binary SDSS J141126.20+200911.1 (SDSS1411). Close WD--BD binaries provide powerful laboratories for studying the effects of stellar irradiation on substellar atmospheres because of their high companion-to-host flux ratios, short orbital periods, and intense UV irradiation from the WD. We measure strong wavelength-dependent variability, with day--night relative flux variations reaching $\sim$280$\%$. The SDSS1411B phase curves are well described by a second-order Fourier model, and correlations among the Fourier amplitudes suggest larger flux and temperature variations in the upper atmosphere. Non-irradiated forward models under chemical equilibrium do not reproduce the observed phase-resolved SDSS1411B spectra, indicating substantial departures from isolated-BD atmospheric structures. In contrast, atmospheric retrievals successfully reproduce both the dayside and nightside spectra. The retrievals reveal a prominent dayside temperature inversion, favor cloud-free atmospheres, and provide evidence for CO$_2$ disequilibrium chemistry. An energy-budget analysis indicates low day-to-night heat redistribution efficiency and constrains the BD cooling age to $\sim$3~Gyr, while the inferred BD radius is significantly more compact than predicted by BD evolutionary models. These results reveal the rich atmospheric dynamics and disequilibrium chemistry of a UV-irradiated BD and demonstrate the power of full-orbit JWST phase-resolved spectroscopy for constraining the atmospheric physics of strongly irradiated substellar objects.

\end{abstract}


\section{Introduction}\label{sec:intro} 

Brown dwarfs (BDs) offer insights into ultracool atmospheres (T$<$2,000~K) without the contrast challenges posed by the host stars of exoplanets. The more favorable observing conditions often allow efficient acquisition of higher-quality data than what is possible for exoplanets. Time-resolved observations probe the atmospheres in different rotational phases, thereby providing important spatial constraints on the atmospheres \citep[e.g.,][]{Buenzli14,Metchev15,Karalidi2016,Biller24,Fuda24,McCarthy25}. These include constraints on condensate cloud distribution \citep[e.g.,][]{Apai13}, pressure-dependent cloud structures \citep[][]{Buenzli12,Yang16,Vos23,Biller24,Miles26}, particle sizes \citep[e.g.,][]{Lew16,Vos23}, and non-equilibrium chemistry \citep[][]{McCarthy25,Miles26}. Multi-epoch high-precision monitoring also revealed that cloud cover evolves rapidly, even on time scales as short as a few hours \citep[e.g.,][]{Artigau09,Gillon13,Apai17}. It was shown that condensate clouds in L/T transition brown dwarfs often trace atmospheric zonal circulation and are shaped by planetary-scale waves \citep[e.g.,][]{Apai17,Apai21,Zhou22,Fuda24}.  Longer-timescale observations also identified spectrally and dynamically distinct regions at high latitudes, evidence for the presence of polar vortices \citep[][]{Apai21,Fuda24b}.

However, the majority of the BDs studied differ from giant exoplanets in that they are not irradiated (or at negligible levels). It is, then, not trivial to transfer the abundant knowledge gained from non-irradiated BDs to irradiated giant exoplanets. 

This difference in irradiation is important. Stellar irradiation impacts atmospheric physics and chemistry \citep[e.g.,][]{Showman20}. In giant planets, external irradiation influences at least three key atmospheric processes: (1) Energy and material transport (longitudinally and vertically), (2) Cloud and haze formation and dispersal, and (3) Photochemistry (e.g., \citealt{Tan20, Parmentier13, Parmentier21, Steinrueck23}).

Close white dwarf-brown dwarf systems (WD-BDs) bridge the gap between non-irradiated BDs and irradiated exoplanets \citep{Showman16,Showman20}: WD-BD systems provide excellent laboratories for studying the effects of stellar irradiation on substellar atmospheres (e.g., \citealt{Littlefair+14, Amaro24, French+24}). 

Compared with hot Jupiters, WD--BD binaries can produce BD infrared phase curves with signal-to-noise ratios up to $\sim10\times$ higher because of the small radii and blue spectral energy distributions of hot WDs. Their orbital periods are also typically $\sim10$--20$\times$ shorter than those of hot Jupiters, allowing complete phase-curve observations through the expenditure of only modest JWST observing time \citep{Amaro24}. In addition, WD emission is generally featureless and stable across the near-infrared, aside from systems containing pulsating WDs (e.g., \citealt{Amaro25}), enabling a clean separation of the BD spectrum from the WD contribution.

Hubble Space Telescope (HST) phase-curve observations of irradiated BDs have revealed irradiation-driven atmospheric circulation and disequilibrium chemistry in a unique regime of irradiation and atmospheric conditions (e.g., \citealt{Lew+22, Amaro23, Lothringer24, Amaro25}). JWST phase-resolved spectroscopy of irradiated BDs, with its broad wavelength coverage, high sensitivity, and high temporal cadence, now provides a timely opportunity to characterize these atmospheres and gain new insights into the physics of irradiated substellar atmospheres \citep{BroskiLaing2026}. 

JWST program GO-4967 (PI: Y. Zhou) provides JWST/NIRSpec phase-resolved spectrophotometry of five close WD--BD binaries spanning equilibrium temperatures of $\sim$750--3300 K, enabling systematic studies of irradiated BDs across a broad range of irradiation environments. \citet{BroskiLaing2026} presented the first atmospheric analysis from this program for ZTF0038B, the least strongly irradiated BD in the sample. Through light-curve modeling, they identified asymmetric CO$_2$ band light curves as signatures of irradiation-driven disequilibrium chemistry. Using both self-consistent forward modeling \citep{Marley21, Morley24} and atmosphere retrievals \citep{Molliere19, Nasedkin24}, they further revealed phase-dependent variations in atmospheric structure and molecular abundances and constrained day-night energy redistribution.
\begin{figure*}[t]
    \centering
    \includegraphics[width=1\textwidth]{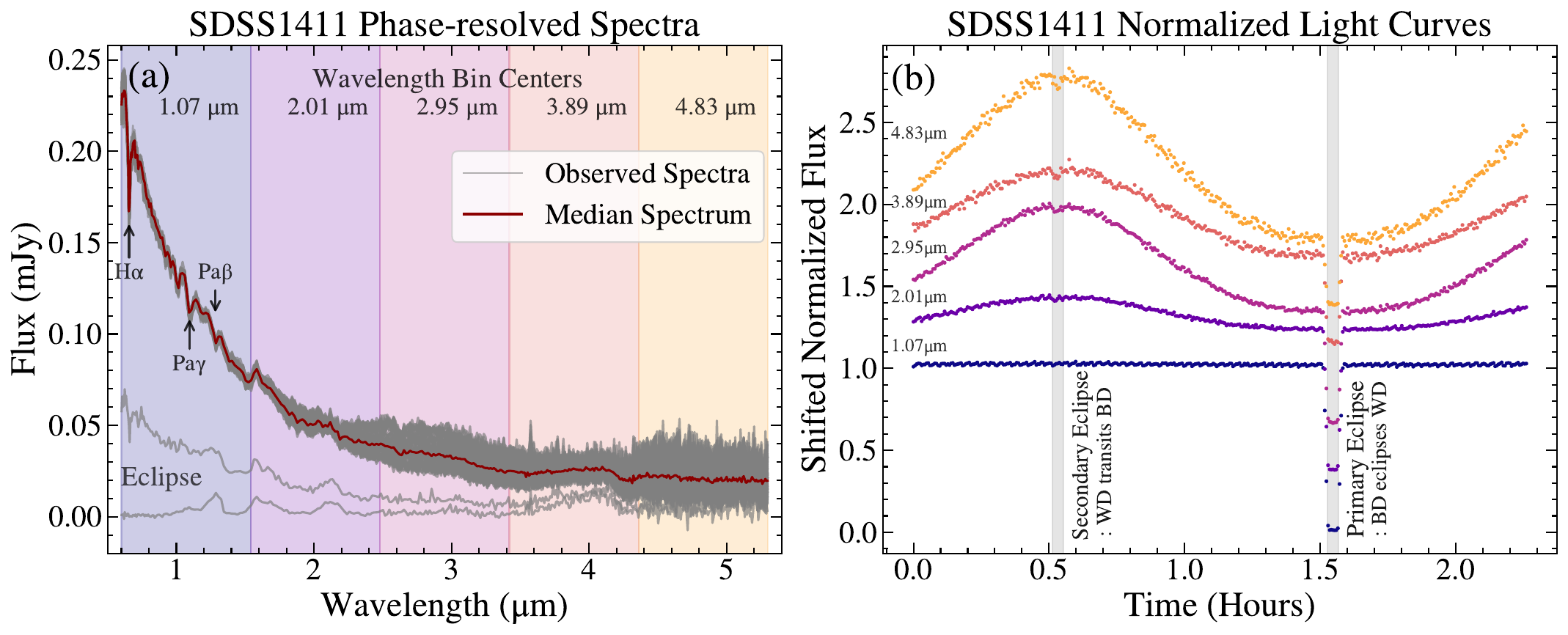}
    \caption{Panel (a): The phase-resolved spectra of the SDSS1411 WD+BD system, shown for every fifth epoch. The shaded regions indicate the wavelength bins used to construct the phase curves shown in panel (b). Panel (b):  Normalized spectrophotometric light curves of the SDSS1411 system. The colors correspond to the shaded wavelength channels in Panel (a). The primary and secondary eclipses occur at elapsed times of $\sim$1.55 and $\sim$0.55 hr, respectively. At wavelengths $\lesssim$2~$\upmu$m, the system light curves are dominated by the stable WD emission, whereas at wavelengths $\gtrsim$2~$\upmu$m, the BD contribution produces strong day--night flux variations.}
    \label{fig:raw_spec_lc}
\end{figure*}

We present an analysis of the second target, SDSSJ141126.20+200911.1 (SDSS1411), which experiences an intermediate level of stellar irradiation (F$\rm{_{irr}}\sim10^{8.6}~erg\,s^{-1}\,cm^{-2}$) within the GO-4967 sample. SDSS1411 was first identified as an eclipsing WD--BD binary by \citet{Beuermann13} in the Catalina Sky Survey \citep{Drake10}. Subsequent ground-based observations constrained the system parameters and basic properties of the BD (SDSS1411B) \citep{Littlefair+14, Casewell15, Casewell18, Yates26}, including a WD effective temperature of $\sim$13,000~K, a BD dayside equilibrium temperature of $\sim1400$~K, a BD mass of $\sim$0.050 M$_\odot$, and an orbital period of $\sim$2.03 hr. Based on these results, \cite{Sainsbury-Martinez21} investigated the atmospheric dynamics of SDSS1411B using 3D global circulation models and suggested that rapid rotation suppresses atmospheric heat transport, particularly downward energy transport into the deep atmosphere, potentially explaining the lack of significant radius inflation. In addition, near-infrared HST phase-curve observations \citep{Lew+22} further characterized day--night variations in the BD atmosphere, including changes in thermal emission and brightness temperature. However, the role of stellar irradiation in shaping the atmospheric thermal structure and disequilibrium chemistry has not yet been investigated in detail using broad-wavelength spectroscopy extending beyond $\gtrsim$2~$\rm \upmu$m.


In this paper, we study the effects of irradiation on the atmosphere of SDSS1411B using JWST/NIRSpec phase-resolved spectroscopy, covering 0.6--5.3~$\upmu$m wavelengths. We analyze the complete phase and multi-wavelength light curve, use atmospheric (forward) modeling, and perform atmospheric retrievals. The paper is organized as follows: Section~\ref{sec:obs} describes the JWST observations and the data reduction process. In Section~\ref{sec:results}, we analyze the phase curves and constrain the binary system parameters. Section~\ref{sec:atm_model} presents the atmospheric modeling and energy transport analyses of SDSS1411B. Finally, we discuss the implications of our results in Section~\ref{sec:dis} and summarize our conclusions in Section~\ref{sec:conclusion}.

\section{Observations and Data Reduction} \label{sec:obs}

\subsection{JWST Observations}
We observed SDSS1411 on 2025 May 27 with JWST/NIRSpec as part of program GO-4967 (PI: Y. Zhou), using the PRISM/CLEAR grating–filter combination with a broad wavelength coverage of 0.6–5.3 $\upmu$m and a low spectral resolving power of R$\sim$100 at 3 $\upmu$m \citep{Birkmann+22, Jakobsen+22}. We performed high-throughput, stable time-resolved spectroscopy in the Bright Object Time Series (BOTS) mode \citep{Espinoza+23} using the 1.6$\arcsec$ × 1.6$\arcsec$ S1600A1 slit without dithering. We covered the full $\sim$2.02 hr orbital period of the SDSS1411 system with 2.27 hr of observations. The time series consists of 75 integrations, each with 120 groups. We used the NRS readout pattern on the SUB512 subarray of the NRS1 detector. 

\begin{figure*}[t]
    \centering
\includegraphics[width=1\textwidth]{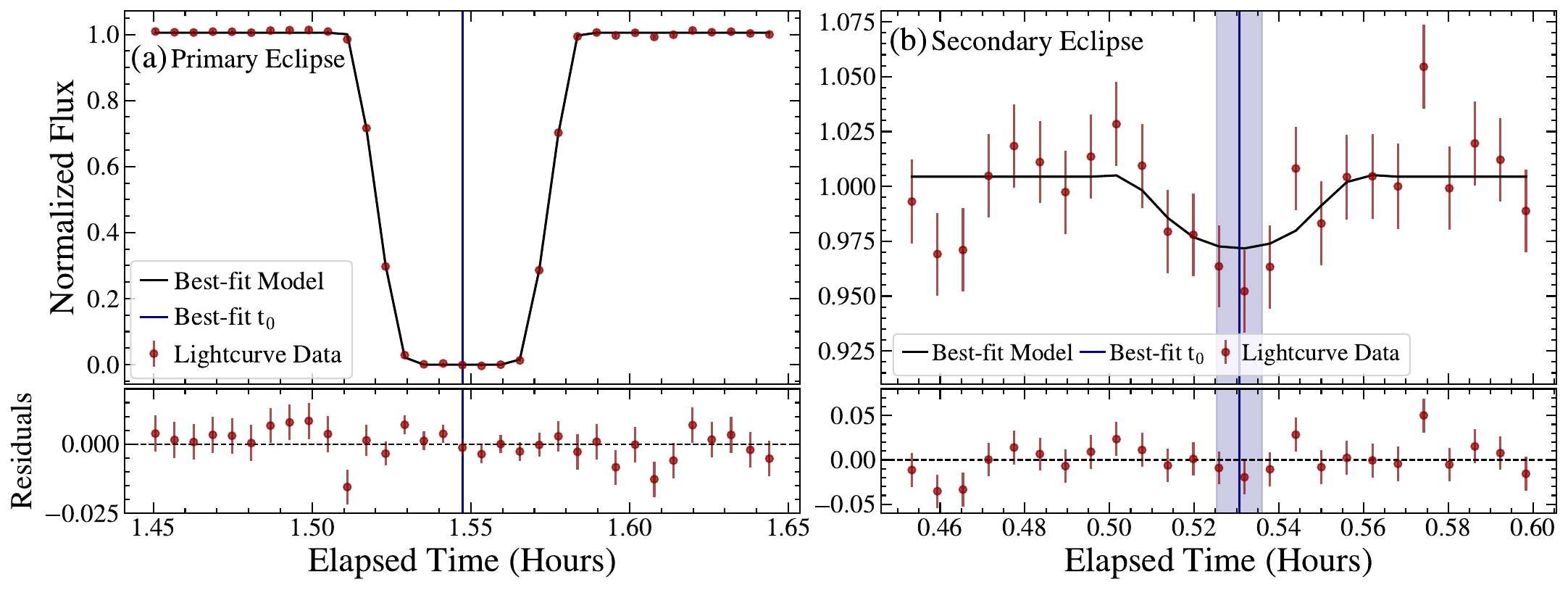}
    \caption{Panel (a): Primary eclipse light curve of SDSS1411 with the best-fit model. The red points show the observed phase curve, and the black solid line represents the best-fit \texttt{batman} transit model. The lower panel shows the residuals between the model and the observed light curve. Panel (b): Same as Panel (a), but for the secondary eclipse.}
    \label{fig:eclipse}
\end{figure*}

\subsection{JWST/NIRSpec Data Reduction}

The JWST/NIRSpec data reduction follows the procedures described in \citet{BroskiLaing2026}, with additional background-subtraction steps summarized below. To extract flux-calibrated time-series spectra, we reduced the JWST/NIRSpec spectroscopic phase curve data of SDSS1411 using the \texttt{Eureka!} pipeline \citep{Bell+22} and the standard \texttt{jwst} data reduction pipeline \citep{Bushouse+23}. We divided each integration into five subgroups to improve the time resolution of the light curve while maintaining sufficient signal-to-noise ratio. We then generated the high-cadence spectrophotometric light curve using the first three stages of the \texttt{Eureka!} pipeline. We constructed the 2D spectral image by fitting the ramps of nondestructive reads using a maximum-likelihood estimator \citep{Brandt24, Brandt24b}, adopting a jump-detection threshold of 15$\sigma$. We derived 1D spectra by performing background subtraction and optimal spectral extraction \citep{Horne86} using an aperture half-width of 3 pixels and a background half-width of 5 pixels. 

To calibrate the 1D spectra from the \texttt{Eureka!} pipeline, we also generated absolute-flux-calibrated spectra for each integration using the standard \texttt{jwst} pipeline. These 1D spectra were obtained through background subtraction using Stage 3 of \texttt{Eureka!}, because the standard \texttt{jwst} pipeline does not perform background subtraction for BOTS observations, followed by optimal spectral extraction based on Stage 2 of the \texttt{jwst} pipeline. We derived the flux calibration factor for the \texttt{Eureka!}-produced spectra from the ratio of the integration-averaged \texttt{Eureka!} spectra to the corresponding flux-calibrated \texttt{jwst} spectrum. Applying this calibration factor to the high-cadence \texttt{Eureka!} spectra, we constructed the final absolute-flux-calibrated, high-cadence spectrophotometric light curve of SDSS1411. 

\section{Results}\label{sec:results}
\subsection{Lightcurve Analysis}\label{sec:lc}

The reduced phase-resolved spectra of SDSS1411 show strong wavelength-dependent flux variations. Figure~\ref{fig:raw_spec_lc} presents the combined WD+BD spectra along with the corresponding light curves integrated over several representative wavelength intervals. In Figure~\ref{fig:raw_spec_lc}(a), the spectra at wavelengths shorter than $\sim$2 $\upmu$m are dominated by WD emission, exhibiting prominent H$\upalpha$, Pa$\upbeta$, and Pa$\upgamma$ hydrogen absorption features at 0.656, 1.095, and 1.284 $\upmu$m, respectively. At wavelengths longer than $\sim$2~$\upmu$m, the BD emission contributes significantly to the total system flux, and its day--night flux modulation drives the large observed variability amplitudes. We also detect significant flux variations in the observed system spectrum associated with the primary eclipse (BD eclipses the WD) at an elapsed time of $\sim$1.55~hr and the secondary eclipse (WD transits the BD) at $\sim$0.55~hr (Figure \ref{fig:raw_spec_lc}b). The flux decrease during the secondary eclipse is small relative to the phase-curve amplitude because the WD, which is smaller than the BD, obscures only a fraction of the BD's projected disk. In Figure \ref{fig:raw_spec_lc}(b), the light curve integrated at 0.6-1.47 $\upmu$m varies by less than 2$\%$, excluding the eclipse phases. In contrast, the phase curves at wavelengths 4.40-5.20 $\upmu$m show larger variability amplitudes up to $\sim$54$\%$ total variation, possibly driven by rotational/orbital modulation of the BD emission.


We modeled the primary and secondary eclipses using the phase-curve data spanning 0.6--5.3~$\upmu$m. The eclipse fits provide mid-eclipse time measurements, allowing us to determine the robust orbital phase of each integration, and constrain the near-infrared limb-darkening parameters. The remaining system parameters are derived in Section~\ref{sec:sys_param} using the latest values reported in the literature. We computed the eclipse models with the \texttt{batman} transit modeling package \citep{Kreidberg15}. Because \texttt{batman} does not include emission from the secondary companion, we first removed the companion contribution from the spectra used for each eclipse fit. For the primary eclipse, we subtracted the average nightside spectrum, measured during $t=1.535$--1.565~hr. For the secondary eclipse, we subtracted the WD spectrum, which we estimated by subtracting the nightside spectrum from the system spectra adjacent to the primary eclipse at $t=1.45$--1.49 and 1.61--1.65~hr. We then integrated the corrected spectra over 0.6--5.3~$\upmu$m for the primary eclipse and 2.0--5.3~$\upmu$m for the secondary eclipse to construct the corresponding light curves. The eclipse fits were performed over elapsed-time intervals of 1.45--1.65~hr and 0.45--0.60~hr for the primary and secondary eclipses, respectively, assuming that the intrinsic BD and WD flux variations are negligible over these short intervals.

We fitted the \texttt{batman} transit model using the \texttt{emcee} package, which implements the Markov Chain Monte Carlo (MCMC) fitting algorithm \citep{Foreman13}. We used four free parameters: mid-eclipse time t$_0$, the ratio of the BD radius over the WD radius r$_p$/r$_s$, a linear limb-darkening coefficient u, and an offset for the light-curve normalization. To reduce model complexity, we fixed the remaining orbital parameters, orbital period $P$, semi-major axis ratio $a/r_{WD}$, and inclination $i$, to the least-$\chi^{2}$ best-fit values reported by \cite{Lew+22}. Table \ref{tab:eclipse_fit} summarizes the adopted priors for the eclipse fit. To properly account for the uncertainties in the phase curves, we inflated the pipeline-produced uncertainties to achieve a reduced $\chi^2=1$ for each phase curve model fit, following \cite{BroskiLaing2026}. 

Figure~\ref{fig:eclipse} shows the observed eclipse light curves together with the best-fit \texttt{batman} models. The modeled light curves are generally consistent with the observed primary and secondary eclipses. The best-fit eclipse parameters are summarized in Table~\ref{tab:eclipse_fit}. The best-fit primary mid-eclipse time, $t_0 = 1.547445 \pm 0.000035$~hr, corresponds to a barycentric-corrected eclipse time of 2460822.8508442(15) MJD$_{\rm TDB}$. This timing is consistent with the linear ephemeris of \cite{Yates26} at epoch 57,149 within $\sim$0.6~s. The time interval between the primary and secondary mid-eclipse times, $\Delta t_0 = 1.0168 \pm 0.0051$~hr, is also consistent with half of the orbital period within the $1\sigma$ uncertainty.

\begin{table}[t]
\centering
\caption{Priors and Best-Fit Parameters for the Eclipse Fit}
\label{tab:eclipse_fit}
\begin{tabular}{lcc}
\toprule
Parameter & Priors for Eclipse Fit & Best-fit Parameters \\
\midrule
$P$ (days)$^*$        & 2.02864 (fixed) & 2.02864 (fixed) \\
$i$ (deg)$^*$       & 86 (fixed)      & 86 (fixed) \\
$a/r_{WD}$$^*$      & 43 (fixed)      & 43 (fixed) \\
\hline
\multicolumn{3}{l}{\bf{Primary Eclipse}} \\
$t_{0}$  & 1.51 -- 1.58    & $1.547445 \pm 0.000035$ \\
$u$               & $-3$ -- $3$     & $0.560^{+0.075}_{-0.080}$ \\
$r_{BD}/r_{WD}$     & 4 -- 5          & $4.7321 \pm 0.0042$ \\
Offset            & 0.9 -- 1.1      & $1.003 \pm 0.0014$ \\
\hline
\multicolumn{3}{l}{\bf{Secondary Eclipse}} \\
$t_{0}$  & 0.5 -- 0.56    & $0.5306 ^{+0.0057}_{-0.0045}$ \\
$u$               & $-4$ -- $4$     & $1.41^{+0.52}_{-1.78}$ \\
$r_{WD}/r_{BD}$     & 0.1 -- 0.3          & $0.164 ^{+0.027}_{-0.030}$ \\
Offset            & 0.9 -- 1.1      & $1.004 \pm 0.005$ \\
\bottomrule
\end{tabular}
*Adopted best-fit values from \cite{Lew+22} 
\end{table}

\subsection{White Dwarf Spectrum Subtraction}\label{sec:wd}

\begin{figure}[t]
    \centering
    \includegraphics[width=1\linewidth]{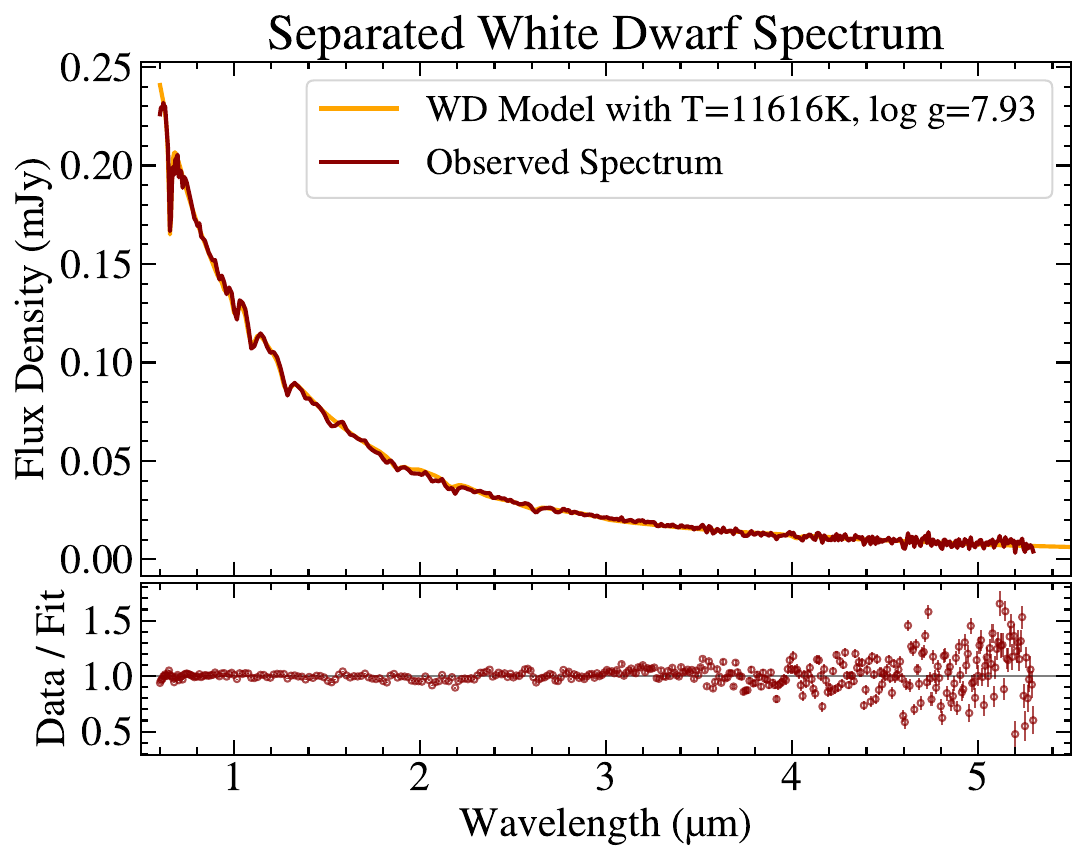}
    \caption{Separated WD spectrum of the SDSS1411 system and the best-fit model WD spectrum using the model grid from \cite{Tremblay13}. The bottom panel shows the ratio between the observed WD spectrum and the best-fit model. The observed spectrum is generally consistent with the model within 2$\pm$14$\%$.}
    \label{fig:wd_fit}
\end{figure}

To construct the phase-resolved spectra of the BD, we separated the WD spectrum from the observed spectra of the unresolved SDSS1411 WD-BD system. We assumed that the combined spectra consist of a stationary WD component and a time-dependent BD component (see Figure \ref{fig:raw_spec_lc} and Section \ref{sec:lc}). Spectra obtained near but outside the primary eclipse contain contributions from both the WD and the BD nightside, whereas the primary-eclipse spectra contain only the BD emission. We therefore extracted the WD spectrum by subtracting the primary-eclipse spectra from the nightside system spectra, based on the assumption that the BD nightside emission remains constant during the night and eclipse phase. We defined the eclipse-phase intervals as $t=1.535$--1.565 ($\phi$ $\sim$ 0-0.008 and 0.994-1), and the night-phase intervals as $t=1.45$--1.49 and 1.61--1.65~hr ($\phi$ $\sim$ 0.03-0.05 and 0.95-0.97). Figure \ref{fig:wd_fit} shows the resulting WD spectrum, whose flux is approximately $\sim$23 times higher than the time-averaged BD flux.


To assess the agreement with theoretical WD models and evaluate the quality of the absolute flux calibration, we compared the separated WD spectrum to pure-hydrogen WD model spectra from \cite{Tremblay13}. We linearly interpolated between the models with $\log g=7.5$ and 8.0 to obtain a model spectrum at $T_{\rm eff}=11,616$ K and $\log g=7.93$, corresponding to the best-fit WD parameters reported by \cite{Vincent24}. To match the spectral resolution, we convolved the model spectra with the dispersion curve of NIRSpec/PRISM. Figure \ref{fig:wd_fit} presents the separated WD spectrum and the best-fit WD model. The fitted scaling factor corresponds to a WD radius of $0.0134 \pm 0.0003~R_{\odot}$, using the Gaia DR3 distance of $177 \pm 4$ pc \citep{gaia23}, consistent with $0.0130 \pm 0.0003~R_{\odot}$ reported by \cite{Lew+22}. The observed and model spectra agree to within relative differences of 2$\pm$14$\%$.  

\subsection{Phase-dependent Brown Dwarf Spectra}\label{sec:fourier}
\begin{figure}
    \centering
    \includegraphics[width=1\linewidth]{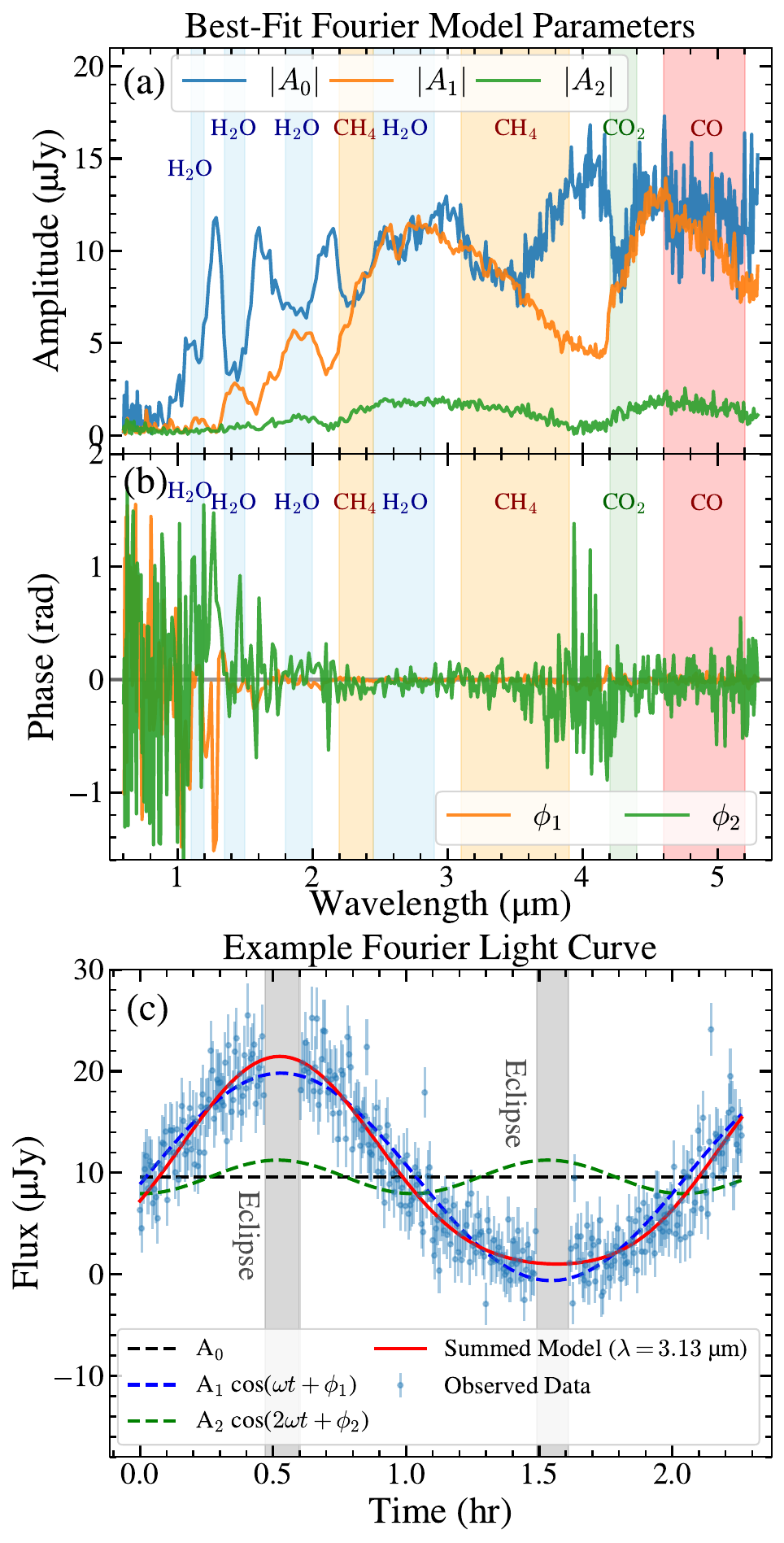}
    \caption{ 
    Panel (a): Amplitudes of the individual Fourier components as a function of the central wavelength of each bin. A$_0$ corresponds to the constant component of the BD emission. A$_1$ and A$_2$ represent the amplitudes of the time-varying BD emission components at each wavelength. Shaded regions indicate wavelengths with strong molecular opacities, which generally coincide with troughs in $\rm |A_0|$ and peaks in $\rm |A_1|$. Panel (b): Same as Panel (a), but for phases of the Fourier components. The best-fit phases are generally consistent with zero phase shift within 1$\sigma$ of the phase distributions. The phase scatter is larger at wavelengths where the Fourier amplitudes are smaller. Panel (c): Example of a second-order Fourier fit to the BD light curve in the wavelength bin centered at 3.13 $\upmu$m. The red line indicates the best-fit model light curve, and colored dashed lines show individual Fourier components. The gray shaded regions mark integrations masked due to eclipse features.}
    \label{fig:fourier}
\end{figure}


\begin{figure*}[t]
    \centering
    \includegraphics[width=1\textwidth]{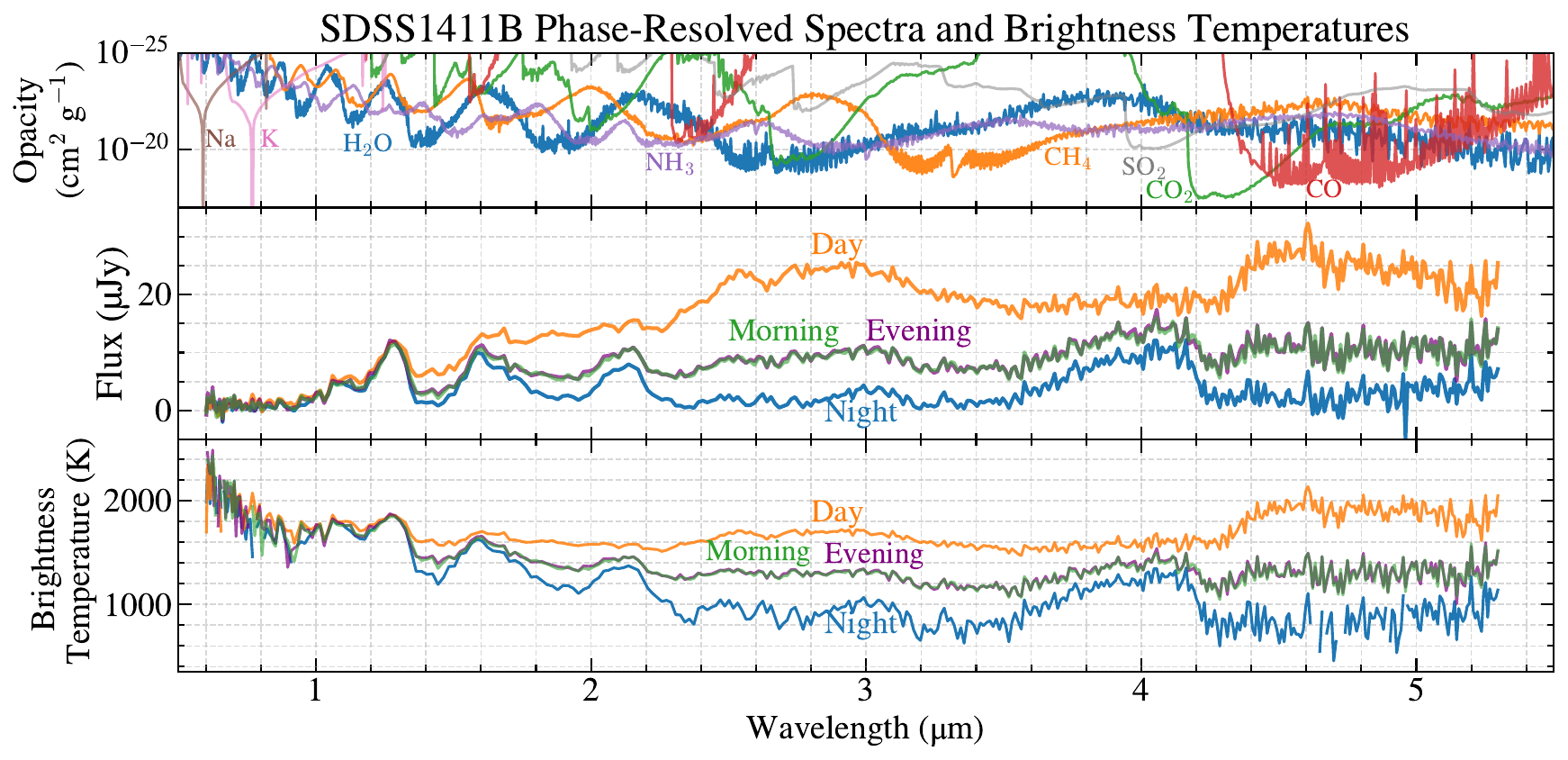}
    \caption{Top: Opacities of common absorbers in exoplanet atmospheres, calculated at a pressure of 0.1 mbar and a temperature of 1,000 K. The y-axis is inverted to better compare the wavelength-dependent opacity trends with the phase-resolved BD spectra shown in the middle panel. Middle: BD spectra at different orbital phases reconstructed using the best-fit Fourier model (Equation \ref{eq:fourier}). Different colors indicate orbital phase: orange, green, purple, and blue correspond to the Day ($\phi = 0$), Morning and Evening ($\phi = \pi/2,~ 3\pi/2$), and Night ($\phi = \pi$) phases, respectively. The phase-resolved spectra show wavelength-dependent flux variation, with an average variation of $\sim$450$\%$ relative to the nightside spectrum. Bottom: Brightness temperatures of SDSS1411B with different orbital phases. The brightness temperatures span $\sim$600-2000 K, with an average day-night temperature variation of $\sim$580~K. }
    \label{fig:phase_opacity}
\end{figure*}

\begin{figure*}[t]
    \centering
    \includegraphics[width=\linewidth]{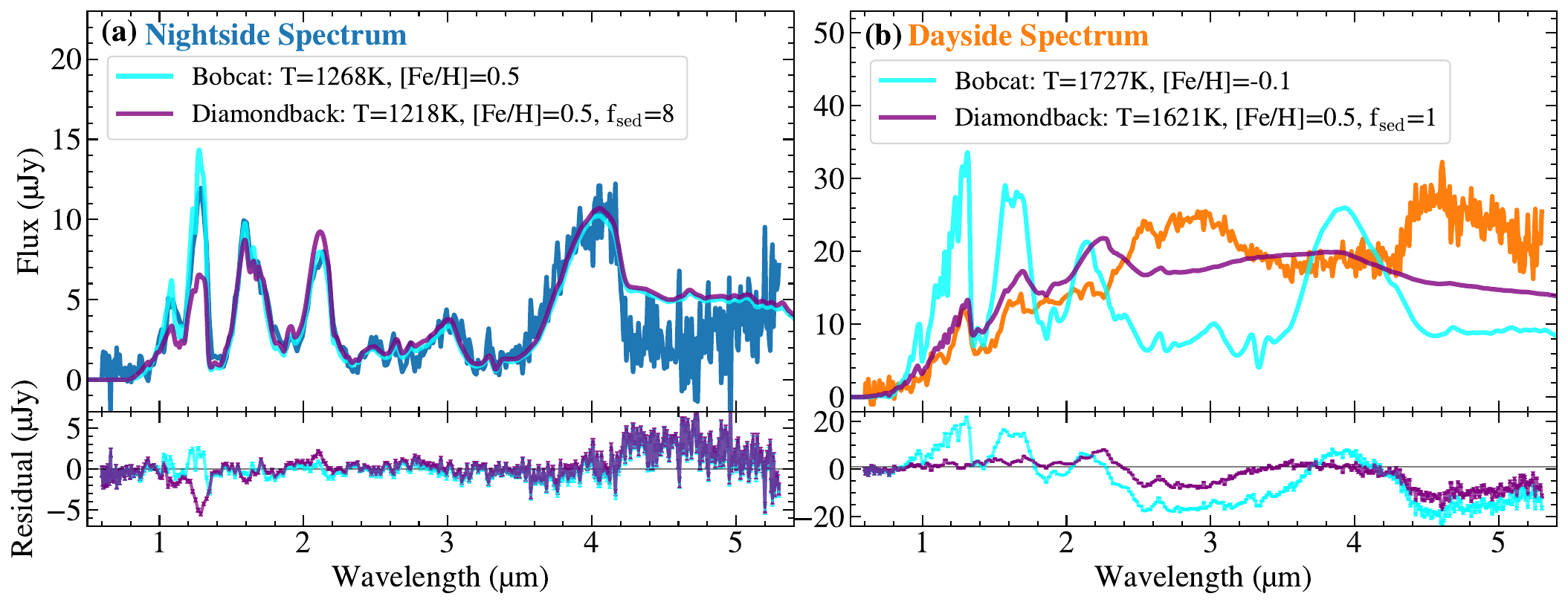}
    \caption{Nightside (Panel a) and dayside (Panel b) BD spectra compared with the best-fit cloudless Sonora Bobcat model (purple) and cloudy Sonora Diamondback model (cyan). The lower panels show the residuals between the observed BD spectra and the best-fit models. The discrepancies between the observed spectra and the equilibrium forward models may reflect departures from equilibrium atmospheric structures driven by strong external WD irradiation.}
    \label{fig:phase_sonora}
\end{figure*}

After subtracting the phase-independent WD component, we derived the phase-resolved BD spectra by using a Fourier model to describe the BD emission modulation (e.g., \citealt{Lew+22, BroskiLaing2026}). The Fourier modeling produces robust phase-resolved BD spectra based on the spectrophotometric data spanning a broad range of orbital phases. We masked the eclipse phases to prevent fitting the eclipse features as intrinsic BD flux variations. We then modeled the phase curve in each wavelength bin using a second-order Fourier series, 
\begin{equation}\label{eq:fourier}
F(\lambda, t) = A_0(\lambda) + A_1(\lambda) \cos(\omega t +\phi_1) + A_2(\lambda) \cos(2\omega t + \phi_2).
\end{equation}
We fixed $\omega$ using the orbital period of 2.028786 hr reported by \cite{Yates26} and defined $t$ as the elapsed time from the mid-eclipse time, $t_0=1.547445$, measured in Section~\ref{sec:lc}. We then fitted the amplitudes and phases of Fourier components in each wavelength bin. Similar to the phase-curve modeling described in Section~\ref{sec:lc}, we inflated the pipeline-produced uncertainties such that the reduced $\chi^2$ of each Fourier model fit is unity, resulting in an average error-inflation factor of $\sim$1.86.

Figure \ref{fig:fourier}(a) shows the best-fit Fourier amplitudes at each wavelength bin. A$_0$ represents the constant (time-independent) component of the BD emission, while A$_1$ and A$_2$ quantify the temporal variation amplitudes of the BD emission at each wavelength. The wavelengths with strong molecular absorption generally correspond to the trough of $\left| A_0\right|$ and peak of $\left| A_1\right|$, implying the relation between photospheric pressures and flux variation amplitudes. Figure~\ref{fig:fourier}(b) shows the best-fit phases for the corresponding Fourier components. The fitted phases are generally consistent with zero, with $\phi_1=-0.01\pm0.30$ and $\phi_2=-0.03\pm0.44$. Larger phase scatter is observed at wavelengths where the corresponding Fourier amplitudes are smaller. The first- and second-order Fourier components show a phase difference of $\phi_1 - \phi_2 \sim$0, indicating the symmetry of the light curve.

We reconstructed phase-resolved BD spectra using the best-fit Fourier model. We estimated the flux density uncertainties by propagating the uncertainties of the best-fit model parameters. Figure \ref{fig:phase_opacity} presents the reconstructed BD spectra at different orbital phases. The integrated dayside BD flux is approximately 280$\%$ brighter than the nightside BD flux. The BD flux density varies by $\lesssim$ 30$\%$ at $\lesssim$ 1.3 and $\sim$ 1.6 $\upmu$m, with the largest variation of $\gtrsim$ 2,000$\%$ at $\sim$ 3.4 and 5 $\upmu$m. The troughs in the BD nightside spectrum generally coincide with opacity peaks of common atmospheric absorbers, including H$_2$O, CO$_2$, and CO. In particular, H$_2$O, CH$_4$, and CO exhibit strong opacity at $\sim$3 and 5 $\upmu$m, which correspond to the spectral regions with the largest variability (see Section \ref{sec:dis_bd}). These regions show prominent absorption features on the nightside, but strong emission features on the dayside, possibly indicating a significant temperature inversion in the dayside atmosphere.

To study the thermal structure of SDSS1411B, we calculated the brightness temperature at each wavelength, defined as the temperature of a blackbody that produces the same specific intensity as the observed flux. The brightness temperature provides an approximate measure of the atmospheric temperature at the layer probed by each wavelength. When combined with a temperature--pressure (T--P) profile without a temperature inversion, the brightness temperature also provides a useful proxy for the relative pressure levels probed at different wavelengths. We derived the brightness temperatures using the Planck function, adopting published values for the BD radius and system distance of 0.70 $\pm$ 0.04 R$_{\rm Jup}$ and 177 $\pm$ 4 pc, respectively \citep{Littlefair+14, Luri18, gaia23}. 

Figure \ref{fig:phase_opacity}(c) shows the brightness temperature as a function of wavelength at different orbital phases. The brightness temperature has strong wavelength dependence, spanning $\sim$600-2,000 K. The average day–night temperature difference is $\sim$610 K. At $\sim$ 3 and 5 $\upmu$m, where the nightside brightness temperature is lowest, the temperature variation is largest, reaching $\sim$1,000 K. In contrast, spectral regions at $\lambda$ $\lesssim$ 1.3, $\sim$1.6, 2.1, and 4 $\upmu$m show much smaller temperature variations of $\lesssim$ 400 K.

\section{Modeling Day-Night Spectral Variation}\label{sec:atm_model}
We investigated the atmospheric structure of SDSS1411B by comparing the observed phase-resolved spectra to atmospheric model spectra. We first updated the system parameters required for the atmospheric modeling using the latest values reported in the literature in Section~\ref{sec:sys_param}. We then performed one-dimensional (1D) forward modeling in Section~\ref{sec:forward}, followed by atmospheric retrieval analysis in Section~\ref{sec:retrieval}. Finally, based on the spectral modeling results, we investigated the energy balance between the day- and nightside hemispheres of SDSS1411B in Section~\ref{sec:e_transport}.

\subsection{Updating System Parameters}\label{sec:sys_param}
We adopted or derived the system parameters required for the atmospheric modeling using measurements reported in previous studies. We first adopted the Gaia DR3 distance of $177\pm4$~pc \citep{gaia23}. We then derived a BD mass of $54.7\pm2.1~M_{\rm J}$ using the binary mass function,
\begin{equation}
\frac{P K_1^3}{2\pi G}
=
\frac{M_{\rm BD}^3\sin^3 i}
{(M_{\rm WD}+M_{\rm BD})^2},
\label{eq:mass_function}
\end{equation}
adopting a WD mass of $M_{\rm WD}=0.562\pm0.028~M_{\odot}$ from \cite{Vincent24}, a radial-velocity amplitude of $K_1=35.0\pm0.7$~km~s$^{-1}$ and orbital inclination of $i=86.5\pm0.4^\circ$ from \cite{Littlefair+14}, and an orbital period of $P=0.08453275006(4)$~day from \cite{Yates26}. We then estimated the orbital separation to be $a=0.69\pm0.01~R_{\odot}$ by applying Kepler's third law using the adopted orbital period and the WD and BD masses. Combining this separation with the eclipse-fit constraint $R_{\rm BD}/a=0.107\pm0.005$ from \cite{Littlefair+14}, we inferred a BD radius of $R_{\rm BD}=0.717\pm0.035~R_{\rm J}$. Finally, using the derived BD mass and radius, we calculated the surface gravity as $\log g=\log(GM_{\rm BD}/R_{\rm BD}^{2})\approx5.42$, which we fixed in the atmospheric modeling.

\subsection{Forward Modeling}\label{sec:forward}
Based on the derived system parameters, we first compared the observed BD spectra with the Sonora Bobcat models \citep{Marley21}, a 1D cloudless grid assuming rainout chemical equilibrium. The model grid spans $2.5 \leq \log$ g $\leq 5.5$ and $200 \leq T_{\rm eff} \leq 2{,}400$ K. We interpolated the model spectra and fitted the observed spectra using the \texttt{species} package \citep{Stolker20}. We implemented the \texttt{UltraNest} nested sampling algorithm \citep{Skilling04, Buchner21} with 400 live points. We used a uniform prior on the effective temperature of $1{,}000 \leq T_{\rm eff} \leq 2{,}400$ K and metallicity of $-0.5 \leq$ [Fe/H] $\leq 0.5$. 

We also compared the BD spectra with 1D cloudy Sonora Diamondback models \citep{Morley24}, which include effects of refractory cloud species. The model grid spans surface gravities from $\log$ g = 3.5 to 5.5 and effective temperatures from 900 to 2400 K. We adopted the same interpolation and fitting procedure used for the Sonora Bobcat modeling to fit the observed BD spectra.

The best-fit Sonora Bobcat model for the nightside spectrum has $T_{\rm eff}=1268$ K and [Fe/H] = 0.5 (Figure~\ref{fig:phase_sonora}a). The model overpredicts the J-band peak by $\sim$20$\%$ and does not reproduce the positive spectral slope at $\lambda > 4.5$ $\upmu$m. The best-fit Sonora Diamondback model ($T_{\rm eff}=1218$ K, [Fe/H] = 0.5, $f_{\rm sed}=8$) provides similar spectral slope at $\lambda > 4.5$ $\upmu$m, but predicts weaker blue-side peaks within the H$_2$O absorption bands ($\sim$1--2 $\upmu$m) than observed, possibly due to cloud-induced reddening. The discrepancy between the observed nightside spectrum and the cloudy Diamondback model with the least reddening and vertically thinnest clouds ($f_{\rm sed}=8$; see Figure \ref{fig:cloud_abun}b) suggests that the atmospheric conditions of SDSS1411B deviate from equilibrium cloud model predictions (see Section \ref{sec:dis_cloud} for further discussion of cloud abundances).


The Sonora Bobcat model does not provide a satisfactory fit to the dayside spectrum, as expected because the Sonora models do not account for external irradiation. The model does not reproduce both the rising continuum slope and the prominent emission features at $\sim$3 and 5 $\upmu$m, with flux deviations exceeding $\sim$60$\%$ of the observed value. The cloudy Diamondback model produces weaker molecular absorption features, in better qualitative agreement with the observed dayside spectrum. However, the Diamondback model likewise does not reproduce the overall spectral shape, particularly near $\sim$5~$\upmu$m, where the predicted flux differs from the observations by more than $\sim$30$\%$.

\begin{table}[t]
\centering
\caption{Priors for the Joint Retrieval.}
\label{tab:priors}
\begin{tabular}{lcc}
\hline\hline
\textbf{Parameter} & \textbf{Prior} \\
\hline
$\log$ g (cgs)&5.42 (fixed)\\ 
D$\rm{_{BD}}$ (pc)& 177 (fixed) \\ 
R$\rm{_{BD}}$ (R$\rm{_{Jup}}$)& $\mathcal{N}(\mu = 0.70, \sigma = 0.04)$ \\
$T_{\rm bottom}$ (K)& $\mathcal{U}(0,8000)$ \\
$\frac{d\log T}{d\log P}$ & $\mathcal{U}(-1, 1)$\\
$\log P _{quench} $(bar) & $\mathcal{U}(-6, 3)$ \\
$[\mathrm{Fe/H}]$ & $\mathcal{U}(-1, 2)$\\
C/O & $\mathcal{U}(0.05, 1.45)$\\
$\log X_{\rm CO_2, global}$$^*$ & $\mathcal{U}(-10, 0)$\\
$\log X_{\rm CO_2, phase}/X_{\rm CO_2, global}$$^*$ & $\mathcal{U}(-5, 5)$\\
$\log X_{\rm CO, global}$$^*$ & $\mathcal{U}(-10, 0)$\\
$\log X_{\rm CO, phase}/X_{\rm CO, global}$$^*$ & $\mathcal{U}(-5, 5)$\\
$\sigma_{\rm lnorm}$$^{**}$ & $\mathcal{U}(1.05, 3)$\\
$\log K_{zz}$$^{**}$ & $\mathcal{U}(5, 13)$\\
$f_{\rm sed}$$^{**}$ & $\mathcal{U}(0, 10)$\\
$\log X_{0,\rm Fe}$$^{**}$ & $\mathcal{U}(-10, 0)$\\
$\log X_{0,\rm MgSiO_3}$$^{**}$ & $\mathcal{U}(-10, 0)$\\
\hline
\end{tabular}
\vspace{-4mm}
\parbox{0.87\columnwidth}{\footnotesize
$^*$ For free chemistry models\\
$^{**}$ For cloudy models}
\end{table}

\begin{figure*}[t]
    \centering
    \includegraphics[width=\linewidth]{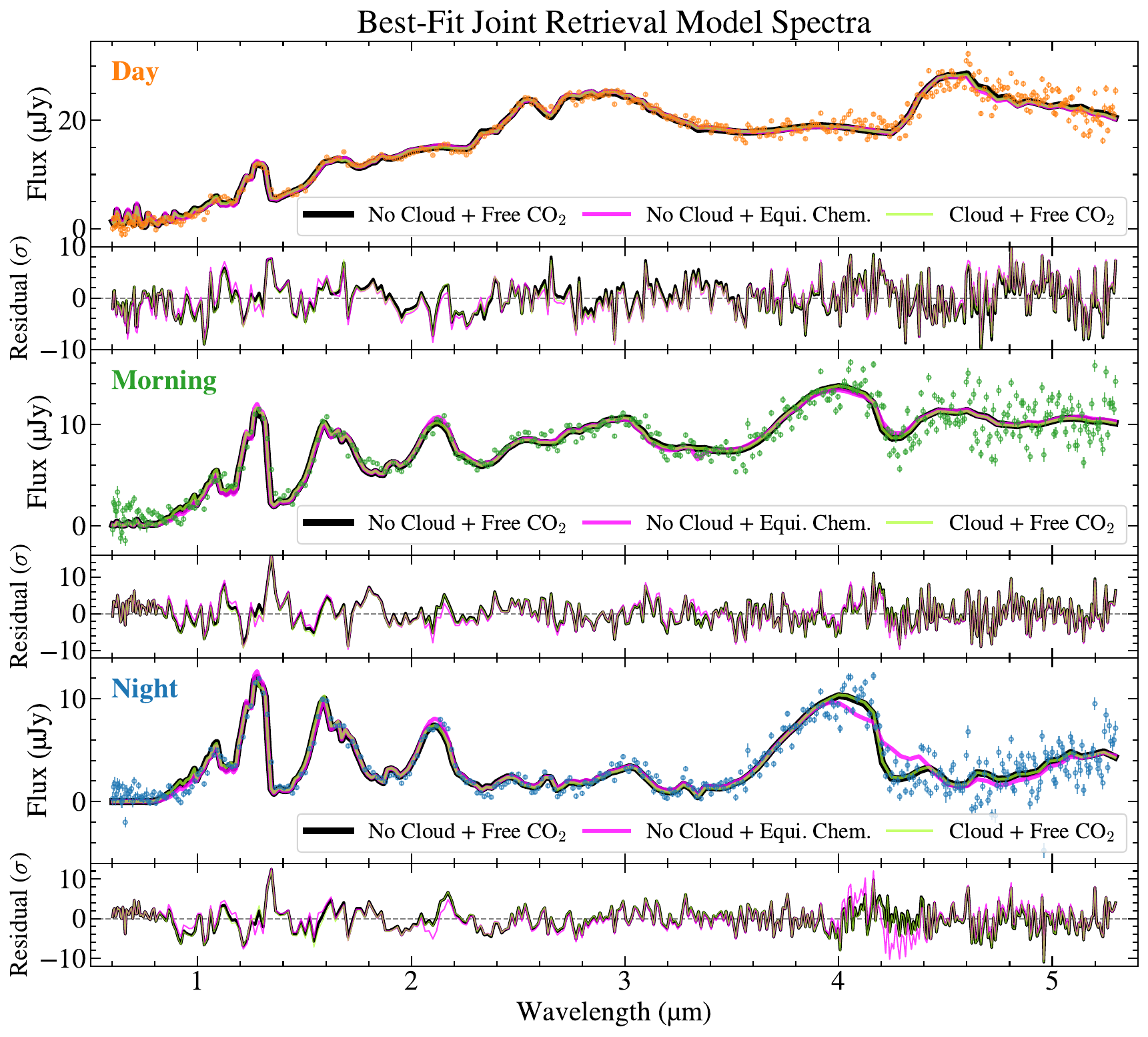}
    \caption{Phase-resolved 
    BD spectra for the Day (orange), Morning (green), and Night (blue) phases, compared with the best-fit retrieval models without clouds and with free CO$_2$ chemistry (black), with clouds and free CO$_2$ chemistry (magenta), and without clouds and with equilibrium chemistry only (lightgreen). The lower panels show the residuals between the observed and model spectra. Relative to the best-fit cloudless model with free CO$_2$ chemistry, including cloud opacity does not significantly improve the spectral fit, whereas the equilibrium-chemistry model does not reproduce the nightside CO$_2$ features near 4~$\upmu$m. }
    \label{fig:retrieval_spec}
\end{figure*}

\begin{table*}[t]
\centering
\caption{Best-Fit Parameters for the Joint Retrieval.}
\label{tab:retrieval_parameters}
\begin{tabular}{lccccccc}
\hline\hline
\multicolumn{8}{l}{\textbf{Global Parameters}} \\
$T_{\rm bottom}$ &
$\left(\frac{d\log T}{d\log P}\right)_1$&
$[\mathrm{Fe/H}]$ &
C/O & R$\rm{_{BD}}$&
&&
\\
\hline
4874$\pm$244&
0.411$\pm$0.028&
0.226$\pm$0.040&
0.512$\pm$0.012&
0.693$\pm$0.004&&&
\\
\hline\hline
\multicolumn{6}{l|}{\textbf{Phase-dependent T--P Profile Properties}} &
\multicolumn{2}{l}{\textbf{Chemistry Parameters}} \\
Phase &
$\left(\frac{d\log T}{d\log P}\right)_2$ &
$\left(\frac{d\log T}{d\log P}\right)_3$ &
$\left(\frac{d\log T}{d\log P}\right)_4$ &
$\left(\frac{d\log T}{d\log P}\right)_5$ &
\multicolumn{1}{c|}{$\left(\frac{d\log T}{d\log P}\right)_6$} &
CO$_2$ &
$\log {\rm P_{quench}}$ \\
\hline
Night &
0.1701$\pm$0.0019&
0.395$\pm$0.055 &
0.50$\pm$0.32 &
-0.01$\pm$0.53 &
\multicolumn{1}{c|}{0.03$\pm$0.54}&
-4.364$\pm$0.067&
0.551$\pm$0.033
\\
Morning&
0.1342$\pm$0.0021&
0.0501$\pm$0.0046&
-0.331$\pm$0.016&
-0.04$\pm$0.14&
\multicolumn{1}{c|}{0.08$\pm$0.54}&
-4.514$\pm$0.134&
-1.19$\pm$0.40\\
Day&
0.1061$\pm$0.0024&
-0.0746$\pm$0.0036&
-0.3773$\pm$0.0093&
0.32$\pm$0.16&
\multicolumn{1}{c|}{-0.02$\pm$0.56}&
-8.77$\pm$1.74&
-3.9$\pm$1.0\\
\hline \hline
\end{tabular}
\end{table*}

\begin{figure}
    \centering
    \includegraphics[width=1\linewidth]{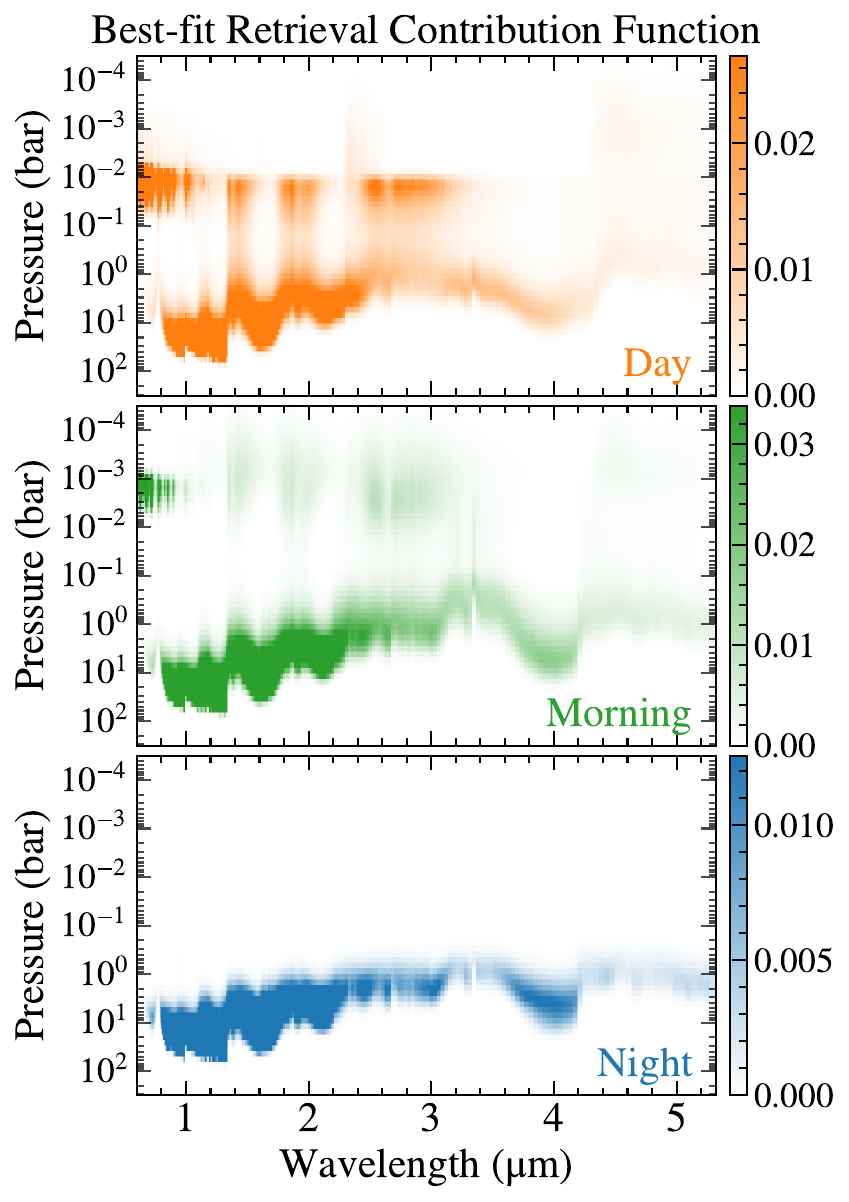}
    \caption{Contribution functions of the best-fit joint retrieval model. Colors indicate the relative normalized contribution of each pressure layer to the observed spectra at the given wavelength. The nightside contribution is concentrated at high pressures of P $\sim$ 1--100~bar, while the dayside and morningside show bimodal contributions from both the deep photosphere (P $\sim$ 1--100~bar) and the inversion region (P$\sim$10$^{-3}$--10$^{-2}$~bar).  }
    \label{fig:2d_contri}
\end{figure}

\begin{figure*}[]
    \centering
    \includegraphics[width=1\textwidth]{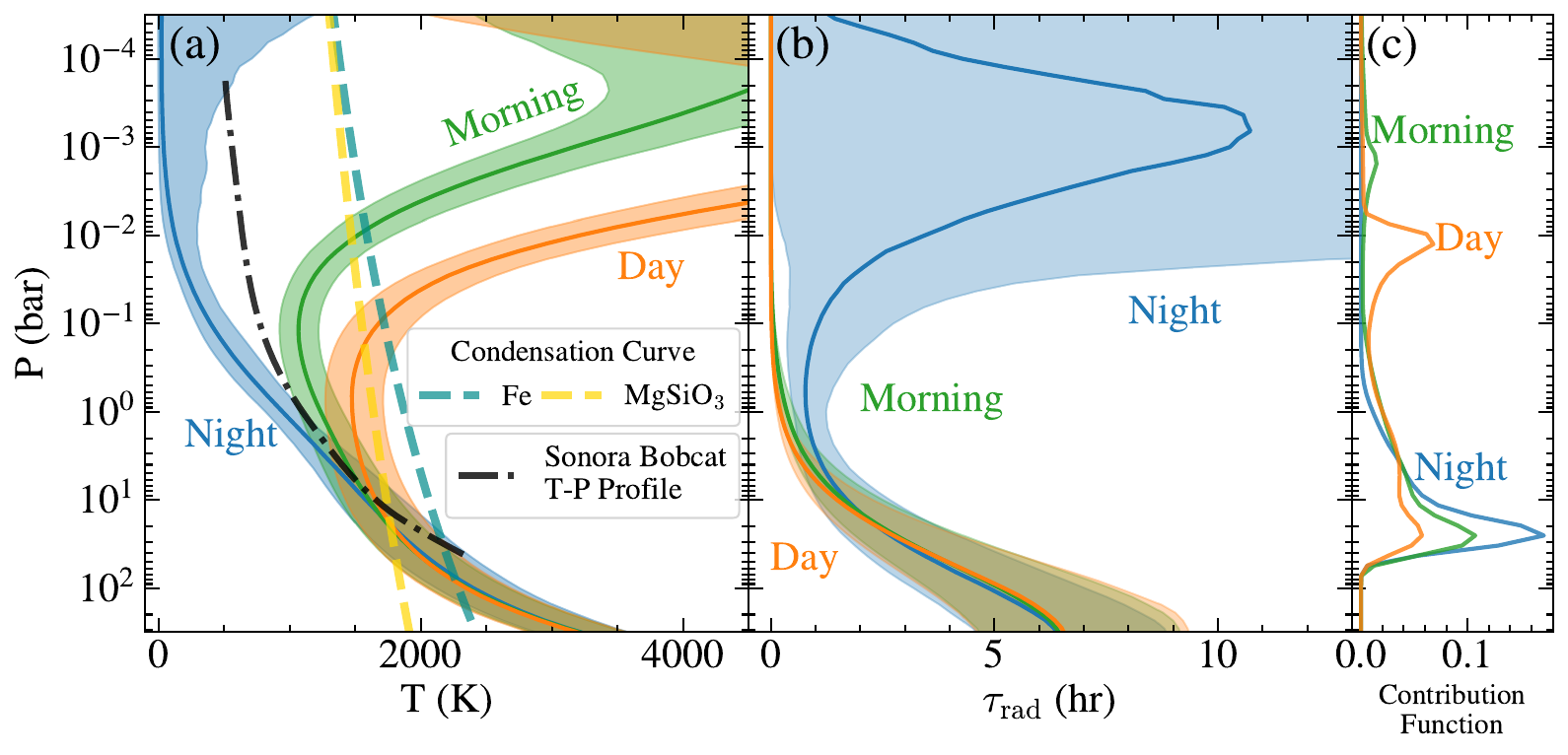}
    \caption{Panel (a): Best-fit T--P profiles for the Day (orange), Morning (green), and Night (blue) phases. The dashed curves show the Fe (cyan) and MgSiO$3$ (yellow) condensation curves, indicating potential cloud formation at $P \sim 10^2$ bar. The black dash-dotted curve shows the Sonora Bobcat model (T${\rm_{eff}}=1276$ K, $\log g=5.42$), which is consistent with the retrieved nightside T--P profile at P $\gtrsim 1$ bar. Panel (b): Radiative timescales derived from the retrieved T--P profiles. At $P \gtrsim 1$ bar, the radiative timescale increases monotonically with pressure. Panel (c): Corresponding contribution functions integrated over 0.6--5.3~$\upmu$m.}
    \label{fig:tp_tau}
\end{figure*}


\subsection{Atmospheric Retrieval Analysis}\label{sec:retrieval}
We further performed multi-phase joint retrievals of SDSS1411B to investigate the detailed atmospheric structures. Because the observed morning- and evening-side spectra are consistent (Figure \ref{fig:phase_opacity}), we considered three representative atmospheric longitudes: nightside ($\phi=0$), morning ($\phi=\pi/2$), and dayside ($\phi=\pi$). Our joint retrieval framework simultaneously analyzes these atmospheric longitudes with multiple 1D models (e.g., \citealt{BroskiLaing2026, Mukherjee26}). We constructed each 1D model using \texttt{petitRADTRANS} (v3; \citealt{Molliere19, Nasedkin24}), which supports a variety of opacity sources and flexible T--P structures. Following the procedures described in \cite{Lothringer25}, we first adopted 80 vertical grid layers evenly spaced in a logarithmic scale of pressure, spanning from 10$^3$ to 10$^{-6}$ bars. We then calculated opacities using tabulated correlated-K tables with the model spectral resolution of $R=100$. Based on the thermal profile construction and opacity calculation, we ran the retrievals with the \texttt{MultiNest} sampling algorithm \citep{Feroz08, Feroz09, Feroz19} using 1,000 live points. The details of modeling the T--P profiles and opacity sources are summarized below.

We modeled the 1D thermal profile of each atmospheric longitude of SDSS1411B using a gradient T--P profile \citep{Zhang23}. We first divided the vertical BD atmosphere into six primary pressure layers evenly spaced in log-pressure. We fitted the temperature at the bottom layer, T$_{bottom}$, and the temperature gradients, d log T / d log P, defined at the boundaries between adjacent layers (namely, 10$^3$, $\sim$16, $\sim$0.3, $\sim$4$\times$10$^{-3}$, $\sim$6$\times$10$^{-5}$, and 10$^{-6}$ bar). A quadratic interpolation was then applied to construct the detailed temperature structure across 80 evenly spaced subgrid layers spanning the six primary layers, given a set of d log T / d log P gradients \citep{Zhang23}. Based on the theoretical prediction from general circulation models (GCMs) \citep{Parmentier13, Lines19}, the phase-dependent T--P profiles were required to converge at the deepest atmosphere by adopting phase-independent parameters for T$_{bottom}$ and d log T / d log P for the deepest layer at P = 10$^3$~bar.  

We then computed the abundances of opacity sources at each atmospheric layer. We included common absorbers in exoplanet and BD atmospheres assuming chemical equilibrium: H$_2$O \citep{Polyansky18}, CH$_4$ \citep{Hargreaves20}, NH$_3$ \citep{Coles19}, HCN \citep{Barber14}, H$_2$S \citep{Azzam16}, FeH \citep{Wende10}, PH$_3$ \citep{Sousa-Silva15}, Na \citep{Allard19}, K \citep{Allard16}, TiO \citep{McKemmish19}, VO \citep{Molliere19}, and SiO \citep{Yurchenko22}. To investigate the disequilibrium processes suggested in \citealt{BroskiLaing2026}, we performed retrievals with free chemistry for CO \citep{Li15} and CO$_2$ \citep{Yurchenko20}, implementing vertically constant abundances. Collision-induced absorption from H$_2$–H$_2$ \citep{Borysow01, Borysow02} and H$_2$–He \citep{Borysow88, Borysow89a, Borysow89b}, and Rayleigh scattering from H$_2$ and He were included as opacity sources \citep{Dalgarno62,Chan65}. We also added H$^{-}$ bound-free and free-free absorption following \cite{Gray08}. Lastly, for cloudy models, we included cloud species Fe and MgSiO$_3$ with crystalline structures, following the cloud model of \cite{Ackerman01}. To avoid interpreting potential scattered or reflected WD irradiation as thermal emission from the BD, we excluded wavelengths shorter than 1~$\upmu$m from the retrieval analysis.

We performed retrievals for absorber abundances both in chemical equilibrium and disequilibrium parameterizations. The absorber abundances in equilibrium were determined by adding [Fe/H] and C/O as free parameters. We accounted for the chemical disequilibrium using the logarithmic quench pressure, log P$_{\rm quench}$ (e.g., \citealt{Zahnle14}). Cloud opacity was parameterized by the width of the cloud particle size distribution, $\rm{\sigma_g}$, the vertical cloud mixing coefficient, $\rm{K_{zz}}$, and the sedimentation efficiency, $\rm{f_{\rm sed}}$ \citep{Ackerman01}. We also determine the condensate abundance at the cloud base using a free parameter $\rm{X_{0}}$ to trace the cloud abundances in each orbital phase. Table~\ref{tab:priors} summarizes the model parameter priors adopted in the retrieval analysis.

We introduced additional flexibility in the chemistry model because initial fits assuming equilibrium chemistry struggled to reproduce the 4.2~$\upmu$m CO$_2$ feature in the nightside spectrum (Figure~\ref{fig:retrieval_spec}; see also the discussion of disequilibrium CO$_2$ chemistry in ZTF0038B in \citealt{BroskiLaing2026}). To assess the robustness of our retrieval setup, we compared models with different treatments of CO and CO$_2$ chemistry, as well as models with and without cloud opacity. Among the tested chemistry schemes, the model with free CO$_2$ retrieval has the highest marginal likelihood ($\ln Z=43965.2$), providing strong evidence \citep{Trotta08} over the free-CO model ($\ln Z=43911.9$), the model with both free CO and CO$_2$ retrievals ($\ln Z=43946.4$), and the model without free retrievals ($\ln Z=43914.4$). We also found that including cloud components does not significantly improve the spectral fit relative to the cloudless model (Figure~\ref{fig:retrieval_spec}), despite the increased model complexity. Instead, the cloudy model is statistically disfavored, providing a lower marginal evidence ($\ln Z = 39452.1$) than the cloudless model ($\ln Z = 43965.2$). We therefore adopt the cloudless retrieval model with free CO$_2$ chemistry as the preferred model for SDSS1411B. The best-fit parameters of the preferred model from the joint retrieval analysis are summarized in Table~\ref{tab:retrieval_parameters}.

The atmospheric retrievals provide generally good fits to the observed multi-phase spectra (Figure~\ref{fig:retrieval_spec}). The retrieved nightside spectrum reproduces the observed positive continuum slope at long wavelengths ($\gtrsim$4~$\upmu$m). Furthermore, the retrieved dayside spectrum successfully reproduces the emission features at $\sim$3 and 5~$\upmu$m, as well as the shallow H$_2$O absorption bands between 1 and 2~$\upmu$m, with relative discrepancies of $\lesssim$20$\%$. 

We plot the wavelength-dependent contribution functions of the best-fit retrieval model in Figure~\ref{fig:2d_contri} to trace the relative contribution of each pressure layer to the observed spectra. At $P\sim1$--100~bar, the photospheric pressures vary inversely with molecular absorption strength (Figure~\ref{fig:phase_opacity}), with stronger absorption features probing lower-pressure layers. The dayside and morningside show additional contributions from low-pressure layers at $P\sim10^{-3}$--$10^{-2}$~bar, possibly probing upper atmospheric layers strongly affected by external irradiation. Our results show that, at wavelengths $\gtrsim$4~$\upmu$m, a relatively broad range of pressures $\lesssim$1~bar contributes to the emerging dayside and morningside spectra, compared to those probed at shorter wavelengths (see also Figures \ref{fig:tp_tau}c and \ref{fig:co2_abun}c).


Figure~\ref{fig:tp_tau}(a) shows the retrieved T--P profiles and corresponding contribution functions, obtained by integrating the wavelength-dependent contribution functions in Figure~\ref{fig:2d_contri}. The nightside exhibits a monotonic thermal structure, whereas the morning and dayside profiles display a prominent temperature inversion at $\sim$0.1 and 1 bar, respectively. The inversion occurs at higher pressure on the dayside than on the morningside, potentially due to stronger WD irradiation on the dayside. The MgSiO$_3$ and Fe condensation curves intersect the retrieved T--P profiles at $\sim$10 and 10$^{2}$ bar, respectively, suggesting cloud formation in the deep atmosphere. At the pressures probed by the nightside observation ($\sim$10 bar), where the nightside contribution function peaks, the retrieved nightside T--P profile is consistent with that of a Sonora Bobcat model with $T{\rm _{eff}}=1268$ K and $\log$ g$=5.42$, adopted from the best-fit model from Section \ref{sec:forward}.  


To investigate the origin of the wavelength-dependent spectral modulations in the phase-resolved BD spectra, we calculated the radiative time constant at each pressure level based on the retrieved T--P profiles. Following \citet{Goody89}, the radiative time constant is given by $\tau = 1/N = c_p P/(4\sigma T^3 g)$, where $N$ is the planetary mean relaxation rate, $c_p$ is the heat capacity at constant pressure, and $g$ is the surface gravity. Figure~\ref{fig:tp_tau}(b) shows the radiative time constant profiles derived from the retrieved T--P profiles. The dayside and morning profiles exhibit increasing radiative time constants with pressure, indicating slower thermal relaxation in deeper atmospheric layers. The nightside profile shows a similar trend at $P \gtrsim 1$ bar, where the contribution function peaks (Figure~\ref{fig:retrieval_spec}). This decreasing trend in timescales at lower pressures indicates a rapid thermal response, large flux and temperature variations in the upper atmosphere. Similar pressure-dependent radiative responses are also found in GCMs of hot Jupiters (e.g., \citealt{Parmentier21}).

\subsection{Energy Transport Balance Analysis}\label{sec:e_transport}

To investigate the energy balance of SDSS1411B based on the retrieved model, we calculated the energy budgets of the nightside and dayside hemispheres under a set of assumptions. We modeled the BD as four equal longitudinal quarters representing the dayside, nightside, morningside, and eveningside, assuming identical spectra for the morning and evening regions. The dayside hemisphere therefore consists of one quarter emitting the dayside spectrum plus one-half of each of the morning- and eveningside quarters, while the nightside hemisphere consists of one quarter emitting the nightside spectrum plus the remaining halves of the morning- and eveningside quarters. We further assumed that the same amount of internal energy is emitted from both hemispheres, that WD irradiation is absorbed only on the dayside hemisphere, and that a fraction of the absorbed energy is redistributed to the nightside. Finally, imposing global energy balance, we solved for the heat-redistribution efficiency, defined as the ratio of the energy transported from the dayside to the nightside to the total absorbed irradiation energy.

\begin{figure}[]
    \centering
\includegraphics[width=1\linewidth]{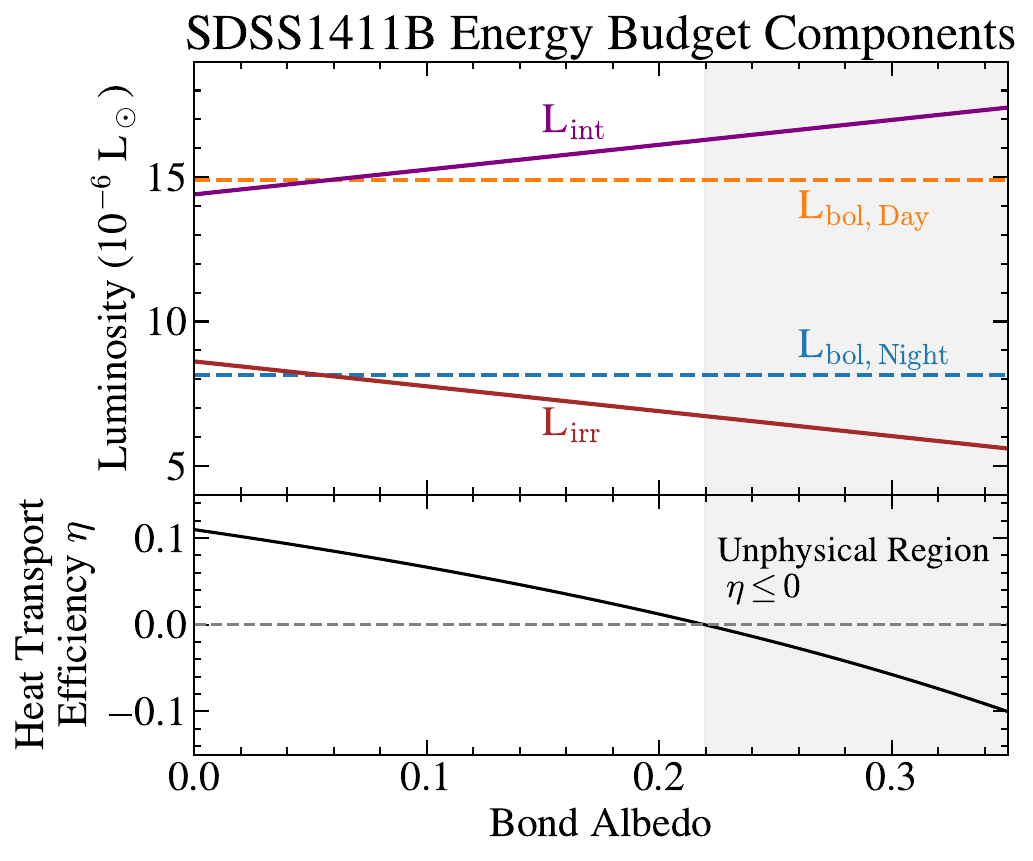}
    \caption{(Top) Energy budget components of SDSS1411B as a function of Bond albedo. (Bottom) Heat transport efficiency between the dayside and nightside hemispheres as a function of Bond albedo. The gray shaded region marks the unphysical regime where the inferred heat transport efficiency becomes negative.}
    \label{fig:e_trans}
\end{figure}

We first constrained the emitted energy from the two hemispheres based on the observed phase-resolved spectra. We derived the bolometric luminosity of each hemisphere by accounting for flux emitted outside the NIRSpec/PRISM wavelength coverage. We performed the bolometric correction of the observed spectra using the best-fit \texttt{petitRADTRANS} retrieval model spectra spanning 0.5–50 $\upmu$m. Based on the ratio between the observed and the model fluxes, we estimated that the observed bandpass (0.6–5.3 $\upmu$m) captures $\sim$84, 92, and 95$\%$ of the BD’s total bolometric flux at day, morning, and night, respectively. The resulting effective temperatures are $1703 \pm$ 58 K, $1446 \pm$ 50 K, and $1272 \pm$ 48 K for the dayside, morningside, and nightside, respectively.

We then constrained the three components of the BD's energy budget: the irradiation, internal, and circulation luminosity. We first calculated the irradiation luminosity (L$_{irr}$), defined as the WD luminosity absorbed by the BD: 
\begin{equation}
    L_{irr} = (1-A) \times F_{inc} \times \pi R^2_{BD}
    \label{eq:L_irr}
\end{equation}
where A is the Bond albedo, F$_{inc}$ is the incident WD flux, and R$_{BD}$ is the BD radius. 
Assuming a global energy balance between the absorbed and emitted energy, the BD internal luminosity is given by
\begin{equation}
    L_{int} = L_{bol, day} + L_{bol, night} - L_{irr}
    \label{eq:L_int}
\end{equation}
where L$_{bol, day}$ and L$_{bol,night}$ are the measured bolometric luminosities of the dayside and nightside hemispheres, respectively. The circulation luminosity ($L_{cir}$), representing the energy transported from the dayside to the nightside, is then
\begin{equation}
    L_{cir} = \frac{L_{int}}{2} +L_{irr} - L_{bol, day}  = L_{bol, night} - \frac{L_{int}}{2}.
    \label{eq:L_cir}
\end{equation}
Finally, we defined the heat transport efficiency, $\eta$ = L$_{cir}$/L$_{irr}$, which quantifies the fraction of absorbed irradiation redistributed from the dayside to the nightside. We note that our simple model does not specifically consider the potential complication that some UV flux is absorbed and thermally re-emitted in the upper atmosphere; this would have an effect similar to a higher Bond Albedo even though it would not be from scattering. We also neglect energy transport through chemical processes, such as photodissociation on the dayside and subsequent recombination on the nightside.


\begin{figure}[]
    \centering
    \includegraphics[width=1\linewidth]{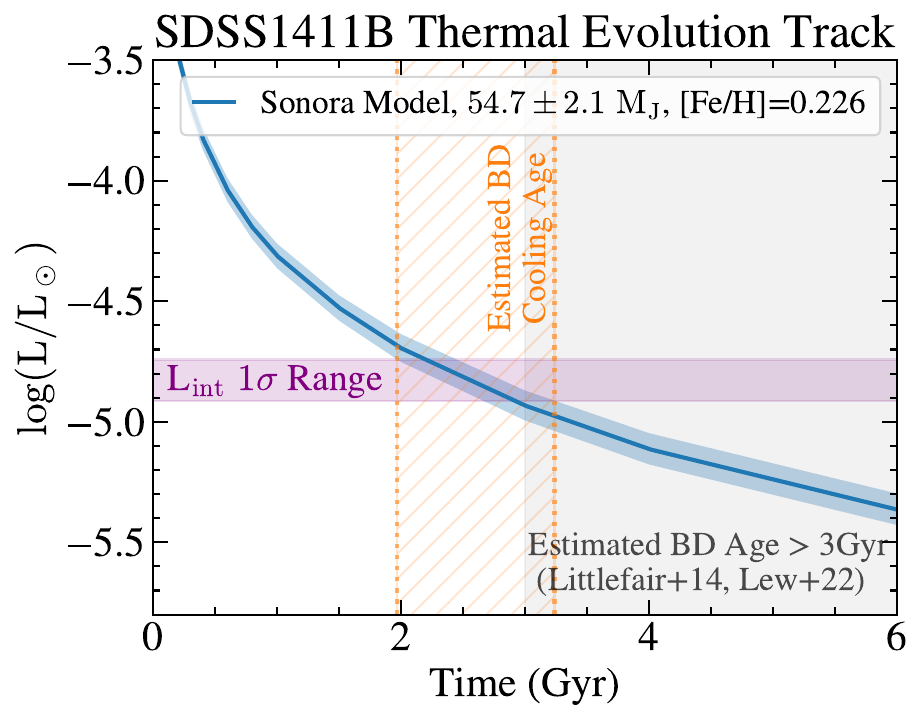}
    \caption{Sonora evolutionary track corresponding to the mass of SDSS1411B, 54.7$\pm$2.1 M$\rm{_J}$. The horizontal shaded region indicates the $1\sigma$ range of $L_{\rm int}$ inferred from the energy transport analysis (see Figure~\ref{fig:e_trans} and Section~\ref{sec:e_transport}). The vertical dotted lines mark the BD cooling-age range corresponding to the overlap between the inferred $L_{\rm int}$ and the Sonora evolutionary track. The gray shaded region marks the age constraint of $\gtrsim$3 Gyr from previous studies \citep{Littlefair+14, Lew+22}, which overlaps with our inferred cooling age of $\sim$3~Gyr.}
    \label{fig:bd_age}
\end{figure}

We obtained each component of the energy budget based on our spectral analysis and system parameters derived in Section \ref{sec:forward}. To estimate the irradiation luminosity from the WD, we adopted a WD effective temperature of $11{,}616\pm299$~K, WD radius of 0.0135$\pm$0.0006 R$_\odot$ \citep{Vincent24}, and orbital separation of 0.69$\pm$0.01 R$_\odot$. We estimated the day- and nightside bolometric luminosities by integrating the best-fit retrieval model spectra over 0.5--50~$\upmu$m and adopting the Gaia DR3 distance of $177\pm4$~pc \citep{gaia23}. We then derived the internal and circulation luminosities based on Equations \ref{eq:L_irr} and \ref{eq:L_cir}.

\begin{figure}[]
    \centering
    \includegraphics[width=1\linewidth]{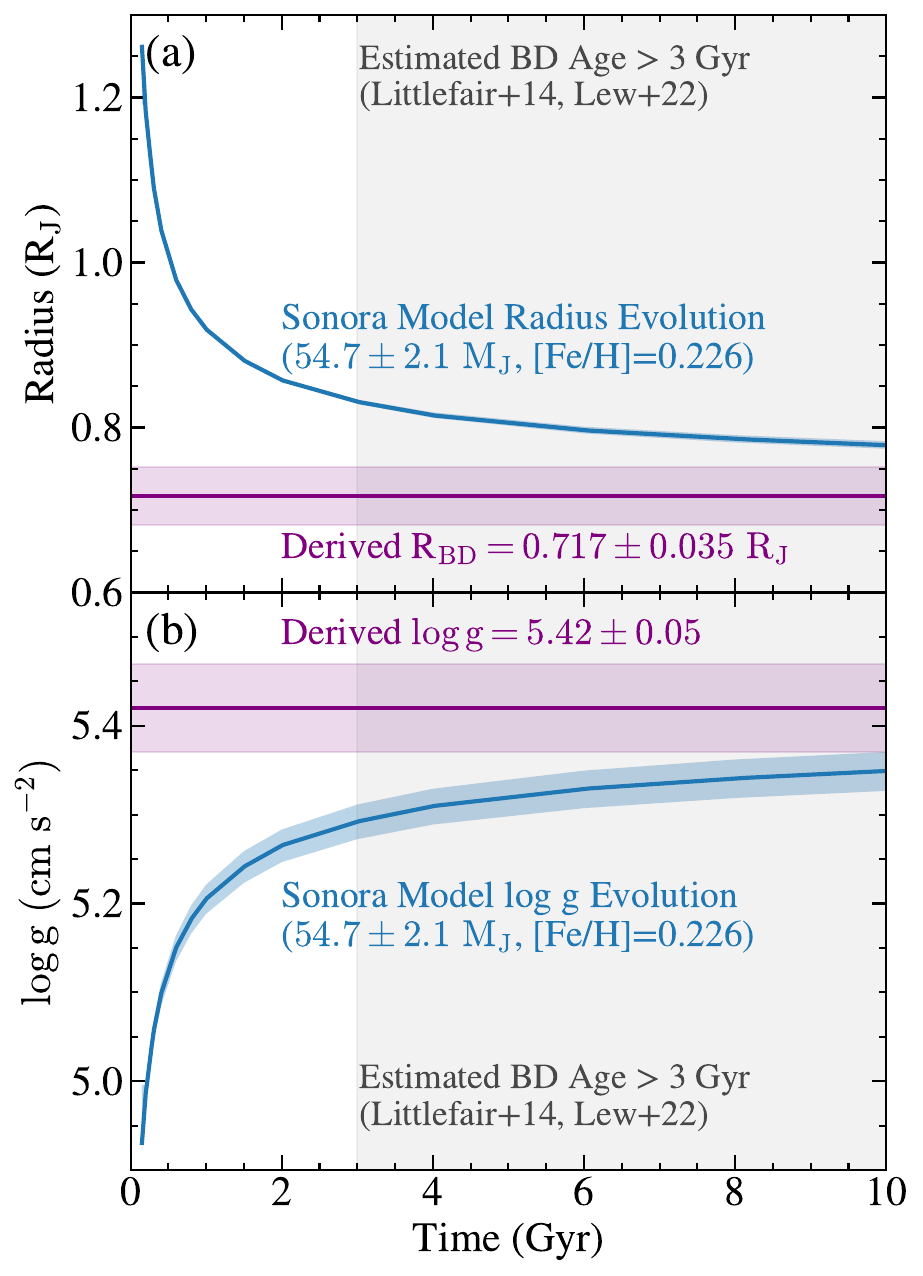}
    \caption{Sonora Bobcat evolutionary tracks for BD radius (Panel a) and $\log g$ (Panel b) corresponding to the mass of SDSS1411B, $54.7\pm2.1~M_{\rm J}$. The horizontal shaded regions indicate the $1\sigma$ ranges of BD radius and $\log g$ derived in Section~\ref{sec:forward}. The inferred radius and $\log g$ do not overlap with the evolutionary model predictions, implying either a very old BD
    age of $\gtrsim10$~Gyr or a more compact BD structure than expected for an isolated BD of the same mass.}
    \label{fig:bd_logg_rad}
\end{figure}

Figure~\ref{fig:e_trans} presents the derived energy budgets as a function of BD Bond albedo. The bolometric luminosity of the dayside hemisphere is $\sim$$1.5 \times 10^{-5}$ $L_{\odot}$, approximately two times larger than that of the nightside hemisphere ($\sim$$8 \times 10^{-6}$ $L_{\odot}$). The absorbed irradiation luminosity decreases linearly with increasing Bond albedo, while the BD internal luminosity increases linearly with Bond albedo. The absorbed stellar irradiation remains comparable in magnitude to the nightside bolometric luminosity. The bottom panel of Figure~\ref{fig:e_trans} shows the heat transport efficiency between the dayside and nightside hemispheres. The maximum efficiency is $\sim$0.1 for zero Bond albedo. The heat transport efficiency decreases with increasing Bond albedo and becomes negative for Bond albedo values greater than $\sim$0.22, providing the upper limit of L$_{\rm int}$ as $\sim 1.6 \times 10^{-5}$ $L_{\odot}$.   

The estimated internal luminosity provides a constraint on the evolutionary age of SDSS1411B. In Figure~\ref{fig:bd_age}, we compare the inferred internal luminosity with thermal evolutionary model tracks. The figure shows the thermal evolutionary track from the Sonora Bobcat evolutionary models \citep{Marley21} for a BD with a mass of $54.7\pm2.1~M_{\rm Jup}$, derived in Section \ref{sec:forward}. We linearly interpolated the evolutionary track with [Fe/H] of 0 and 0.5 to obtain the track with [Fe/H]=0.226, the best-fit metallicity from the retrieval analysis (Table \ref{tab:retrieval_parameters}). The inferred internal luminosities for the range of assumed Bond albedos constrain the BD cooling age to $\sim$2.5~Gyr. Considering the uncertainties in $L_{\rm int}$ and the Sonora model parameters, this estimate is marginally consistent with the $\sim$3--10~Gyr ages inferred from the BD mass--radius relation in previous studies \citep{Littlefair+14, Lew+22}.

We further compared the derived BD surface gravity and radius with predictions from evolutionary models. Figure~\ref{fig:bd_logg_rad} shows the evolution of BD $\log g$ and radius from the Sonora Bobcat models \citep{Marley21}. The derived values of $\log g=5.42\pm0.05$ and $R_{\rm BD}=0.717\pm0.035~R_{\rm J}$ from Section~\ref{sec:forward} correspond to an age older than 10~Gyr, exceeding the upper limit of the model grid. This discrepancy between the age inferred from the internal luminosity and BD compactness may indicate either a more compact structure or a higher internal luminosity than expected for an isolated BD of the same mass.

\section{Discussion}\label{sec:dis}
\subsection{Lightcurve Interpretation}\label{dis:light_curve}

We confirmed significant spectrophotometric variations in the SDSS1411 system (see Figures \ref{fig:raw_spec_lc} and \ref{fig:phase_opacity}), consistent with previous HST and ground-based observations. The BD flux varies by $\sim$48$\%$ from the nightside to the dayside at 1.1–1.6~$\upmu$m, broadly consistent with the $\sim$38$\%$ variability in the HST synthetic photometry by \cite{Lew+22}. We found larger flux variations at longer wavelengths, reaching $\sim$260$\%$ in the $K$ band and $\gtrsim$700$\%$ at $\lambda > 4.5~\upmu$m, which may be attributed to the dayside temperature inversion or UV-induced fluorescence, as suggested by \cite{Casewell15,Casewell18} based on ground-based observations.

The best-fit mid-eclipse time shows no significant phase offset from the trough of the Fourier model of the BD flux variation within the uncertainties ($\phi_1\sim\phi_2\sim0$ in Figure \ref{fig:fourier}, see Section \ref{sec:fourier}). The expected secondary-eclipse phase (orbital phase of 0.5) is also consistent with the peak of the best-fit Fourier model. This near-zero phase offset is consistent with the results of \cite{Lew+22}, suggesting that the hottest region of the BD is centered near the substellar point. The absence of a significant phase offset indicates that the radiative cooling timescale is shorter than the horizontal advection timescale, implying weak or insignificant equatorial (zonal) wind transport (see, e.g., \citealt{Tan20, Lee20}).

\subsection{White Dwarf and System Parameters}

We find that the extracted WD spectrum presented in Section~\ref{sec:wd} generally agrees with the theoretical WD model, with a relative difference of $2\pm14\%$. This agreement supports both the best-fit WD parameters of $T=11{,}616$~K and $\log g=7.93$ adopted from \cite{Vincent24} and the quality of the absolute flux calibration described in Section~\ref{sec:obs}. The WD radius adopted from \cite{Vincent24}, 0.0135$\pm$0.0006 R$_\odot$, is also consistent within 1$\sigma$ uncertainties with the WD radius corresponding to the best-fit scale factor from Section \ref{sec:wd}.

The WD temperature adopted in this study, derived from Gaia low-resolution spectrophotometry, is lower than the $13{,}000\pm300$~K value reported by \cite{Littlefair+14} based on VLT/X-shooter spectroscopy. This discrepancy may arise from the different distance estimates (190~pc from \citealt{Littlefair+14}) and systematic differences among WD parameter-estimation methods. However, the resulting change in the incident WD flux at the BD is only $\sim0.2$~dex, leaving our main conclusions of inefficient heat redistribution and a BD cooling age of $\sim2.5$~Gyr largely unchanged. Moreover, the differences in the adopted WD parameters are not expected to significantly affect our primary conclusions regarding the atmospheric physics of the irradiated BD, which are based directly on the observed BD spectra.

We also find that the observed mid-eclipse time is consistent with the linear ephemeris from \cite{Yates26} at an epoch of $\sim6\times10^4$ orbital cycles. Combined with the mid-eclipse timing data from \cite{Yates26}, we found no significant statistical evidence for orbital evolution over the current observational baseline of $\sim10$ years, despite the very short orbital period of $\sim2$~hr. To further assess the stability of the SDSS1411 system, we calculated the secular evolution of the orbital period for a circular orbit given by \citep{Blanchet14}:
\begin{equation}
    \dot{P} = -\frac{192\pi}{5c^5}\left(\frac{2\pi G}{P}\right)^{5/3}\frac{M_{WD}M_{BD}}{(M_{WD}+M_{BD})^{1/3}} \sim -4.6\times10^{-14}.
    \label{eq:P_evol}
\end{equation}
Substituting the parameters for the SDSS1411 system, we estimate a cumulative phase shift of $\Delta t = \frac{1}{2}\frac{\dot{P}}{P}t_{\rm epoch=57149}^2 \sim 0.5$~s, which would be difficult to robustly detect from a single mid-eclipse measurement. The corresponding orbital decay timescale due solely to gravitational wave emission \citep{Peters64} is
\begin{equation}
T = \frac{5c^5a^4}{256G^3}\frac{1}{M_{\rm WD}M_{\rm BD}(M_{\rm WD}+M_{\rm BD})} \sim 1.9~{\rm Gyr},
\label{eq:T_grav}
\end{equation}
which is comparable to the estimated evolutionary age of SDSS1411B of $\sim2.5$~Gyr. Thus, although gravitational-wave emission should drive secular orbital decay in this short-period and close binary, the expected timing signal over the current observational baseline of $\sim$10~yr is too small to be detected.

\subsection{Phase-dependent Brown Dwarf Spectra}\label{sec:dis_bd}
The Fourier model in Section \ref{sec:fourier} robustly reproduces the observed spectra, and the Fourier amplitudes trace the flux variability of different atmospheric layers. The correlations between the Fourier amplitudes of the stationary ($A_0$) and time-varying components ($A_1$ and $A_2$) reflect the interplay between photospheric pressure at different wavelengths and the corresponding radiative cooling timescale. The opposite wavelength dependence of $|A_0|$ and $|A_1|$ shown in Figure \ref{fig:fourier}(b) indicates that spectral regions with stronger absorption exhibit larger variability amplitudes. Wavelengths with high molecular opacity generally probe higher altitudes in the atmosphere (e.g., \citealt{Yang16}) and correspond to lower photospheric pressures. As shown in Figure \ref{fig:tp_tau}(b), lower pressure is associated with shorter radiative timescales, which tend to indicate larger flux and temperature variations. This naturally explains the observed correlation between molecular absorption strength and variability amplitude and is also consistent with similar pressure-dependent trends in phase-curve offsets found in recent hot-Jupiter GCMs (e.g., \citealt{Coulombe26}).


Brightness temperature measurements provide an additional probe of flux variability across different atmospheric pressure levels. The $\sim$3 and 5~$\mu$m bands exhibit the lowest nightside brightness temperatures (Figure~\ref{fig:phase_opacity}), suggesting that they probe relatively low-pressure layers at $\sim$1 bar given the monotonic nightside T--P profile (Figures~\ref{fig:2d_contri} and \ref{fig:tp_tau}a). These layers have relatively short nightside radiative timescales of $\sim$1 hour (Figure~\ref{fig:tp_tau}b) and are significantly influenced by external irradiation, consistent with the large flux and temperature variations observed at $\sim$3 and 5~$\mu$m. On the other hand, spectral regions with weaker molecular opacity, such as $\lambda\lesssim1.3$, $\sim$1.6, 2.1, and 4~$\upmu$m, probe relatively deeper atmospheric layers and exhibit smaller variations in flux and brightness temperature.


\subsection{Nightside Spectrum: Field Brown Dwarfs and Atmospheric Models}
\begin{figure}[]
    \centering
    \includegraphics[width=1\linewidth]{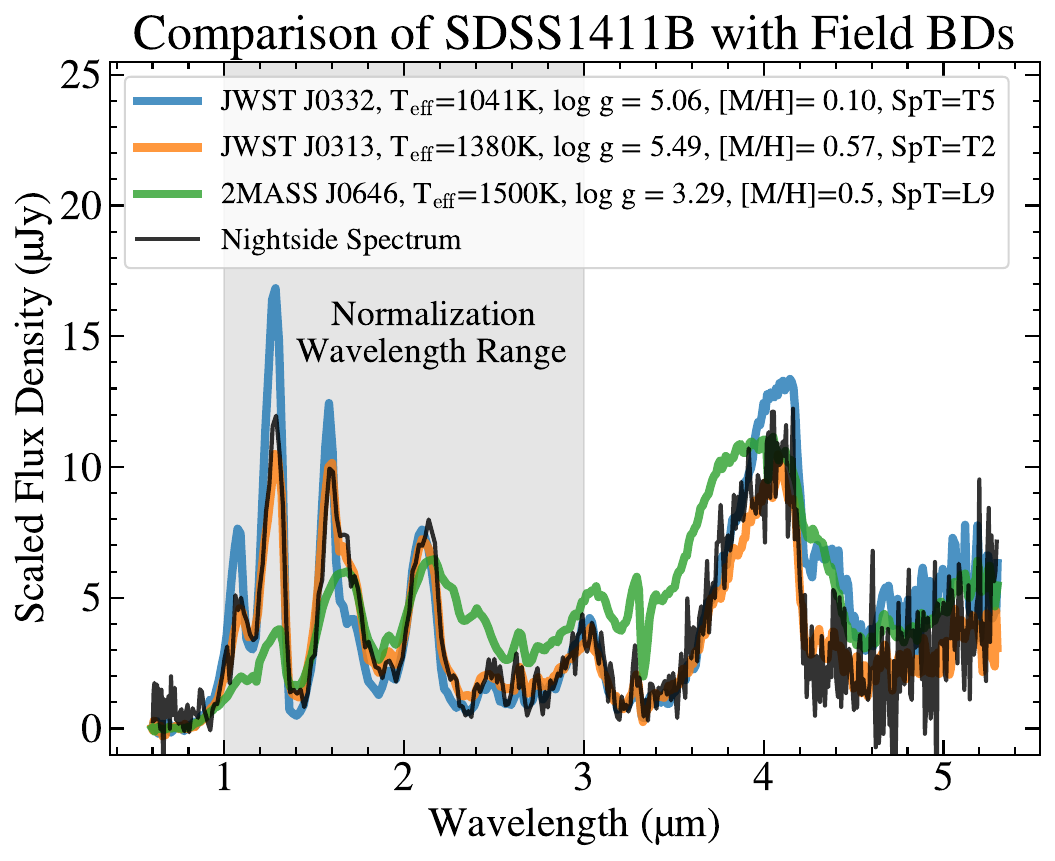}
    \caption{Comparison of the nightside spectrum of SDSS1411B with field BD spectra from \cite{Tu25}. The gray shaded region marks the reference wavelength range used to normalize each spectrum based on its integrated flux. Among the comparison spectra, the T2 dwarf provides the closest match to the observed nightside spectrum of SDSS1411B.}
    \label{fig:field_bd}
\end{figure}

\begin{figure*}[t]
    \centering
    \includegraphics[width=1\textwidth]{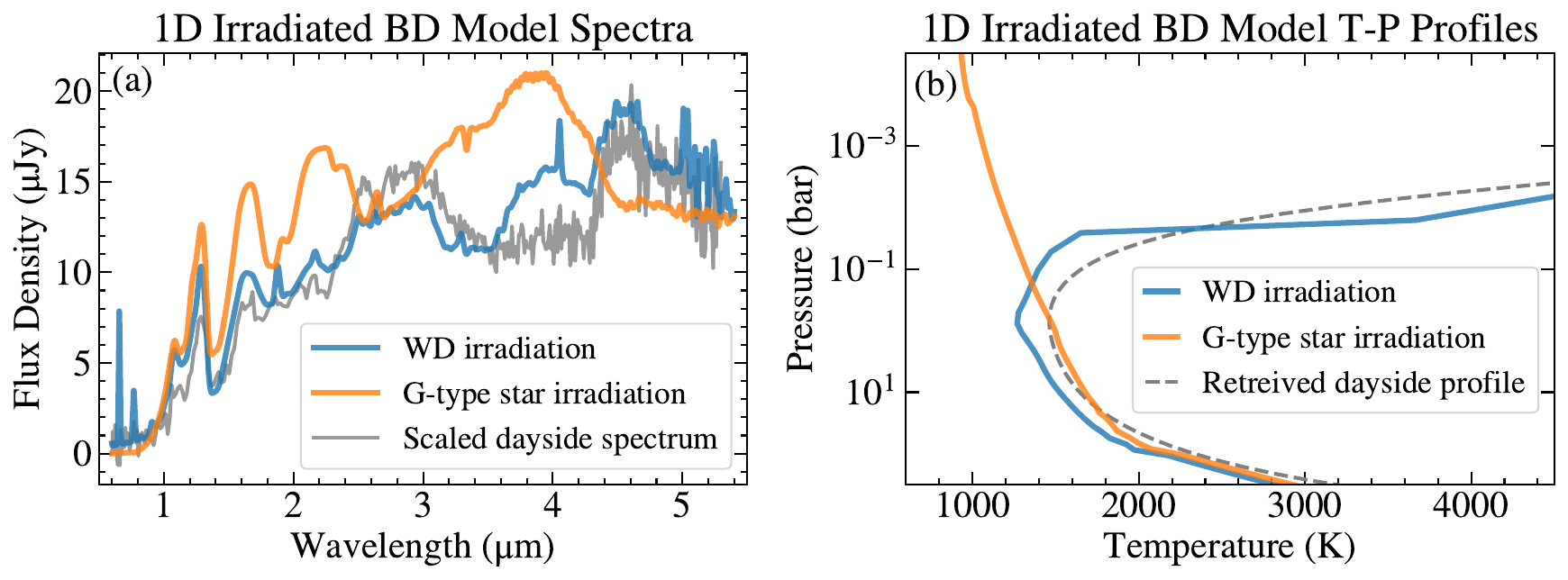}
    \caption{Panel (a): Model spectra from 1D BD atmosphere models irradiated by a WD (blue) and a G-type star (orange), compared with the observed SDSS1411B dayside spectrum (gray). Similar to SDSS1411B, the WD-irradiated model exhibits both absorption bands at short wavelengths and prominent emission features at $\sim$3 and 4.5~$\upmu$m. Panel (b): Corresponding T--P profiles for the two irradiated models, along with the best-fit retrieval profile for the SDSS1411B dayside. Both the WD-irradiated model and the retrieved SDSS1411B dayside profile exhibit a pronounced temperature inversion near 1~bar.}
    \label{fig:1d_model}
\end{figure*}


The forward modeling (Figure~\ref{fig:phase_sonora}a) and retrieval analysis (Figure~\ref{fig:retrieval_spec}) for the observed nightside spectrum provide different atmospheric properties and T--P profiles. The best-fit Sonora Bobcat forward model has $T_{\rm eff}=1268$~K, in good agreement with the retrieved effective temperature of $\sim1271$~K. However, the forward model converges on the highest metallicity value in the model grid ([Fe/H] = 0.5), exceeding the retrieved value of [Fe/H] $\sim0.3$ (Table~\ref{tab:retrieval_parameters}), suggesting departures of the retrieval model from the equilibrium assumptions of the forward model. Furthermore, the retrieved and forward-model T--P profiles are consistent near $P\sim1$ bar but diverge at lower pressures. The steeper retrieved temperature gradient in the upper atmosphere may reflect the effects of external irradiation on the atmospheric structure of SDSS1411B, or systematics of the gradient T--P profile parameterization in pressure regions with small contribution function (Figure~\ref{fig:tp_tau}c).

Surprisingly, the nightside spectrum of the irradiated BD SDSS1411B resembles that of isolated BDs. The retrieved nightside effective temperature of $\sim1270$ K corresponds to an early-T spectral type based on the spectral type--effective temperature relation of \citet{Stephens09, Dupuy17}. Figure~\ref{fig:field_bd} compares the JWST NIRSpec/PRISM spectra of T5, T2, and L9 field dwarfs from \citet{Tu25}. The T2 spectrum closely matches the nightside spectrum of SDSS1411B, reproducing both the H$_2$O absorption bands at $\sim$1--2~$\upmu$m and the CO$_2$ feature at $\sim$4.2~$\upmu$m. In contrast, the T5 and L9 spectra do not reproduce the relative peak heights within the H$_2$O bands and the widths of the spectral features. The agreement between the isolated T2 dwarf spectrum and the SDSS1411B nightside spectrum suggests that stellar irradiation has a limited influence on the nightside atmosphere, likely due to inefficient day-to-night heat transport (see Section~\ref{sec:e_transport}) and the limited impact of common-envelope evolution on the BD spectral features \citep{BroskiLaing2026}.


\subsection{Dayside model fits}

The non-irradiated Sonora forward models do not reproduce the observed dayside emission features at $\sim$3 and 5~$\upmu$m in Figure \ref{fig:phase_sonora}. The lack of irradiation prevents the 1D models from forming any upper-atmosphere temperature inversion \citep{Marley21,Morley24}.

In contrast, the best-fit retrieval model provides a good fit to the dayside spectrum (Figure~\ref{fig:retrieval_spec}). The retrieved dayside T--P profile exhibits a hotter upper atmosphere than the corresponding night- and morningside profiles. The T--P profile also shows a significant temperature inversion at $\sim10^{-1}$~bar, similar to those observed in highly irradiated hot Jupiters (e.g., \citealt{Evans17}), explaining the prominent $\sim2.8\upmu$m H$_2$O and  4.6~$\upmu$m CO emission features. However, unlike any previous spectra of irradiated planets or BDs with inversions, SDSS1411B also exhibits molecular features in absorption, namely, H$_2$O at 1.4~$\upmu$m. This can only be explained by a temperature inversion that exists at pressures probed by the 2.8~$\upmu$m H$_2$O band but not at pressures probed by the 1.4~$\upmu$m band, placing a strict constraint on the shape of the T--P profile. Interestingly, the apparent absence of strong absorption or emission from the 1.9~$\upmu$m H$_2$O band is consistent with it probing the isothermal ``transition" region of the inversion.

While this behavior has never before been seen in an atmosphere, such a scenario for CO was hypothesized to explain the low K-band eclipse depth of WASP-18b compared to the high \textit{Spitzer} 4.5~$\upmu$m eclipse depth \citep{Sheppard17}. However, subsequent observations of WASP-18b with JWST/NIRISS/SOSS showed no evidence for a low eclipse depth at 2~$\upmu$m and were instead consistent with both fundamental and overtone CO bands in emission \citep{Coulombe23}.

SDSS1411B represents the lowest-temperature strongly irradiated substellar object known to exhibit a temperature inversion. Previous theoretical modeling has identified objects above irradiation temperatures of about 2000~K as being susceptible to temperature inversion due to the absorption of the intense incoming flux by short-wavelength absorbers like TiO, VO, Fe, and other metal species, present only in the gas phase for temperatures above refractory condensation temperatures \citep{Hubeny03,Fortney08,Lothringer18,Gandhi19}. With an equilibrium temperature around 1600~K, we expect the canonical optical absorbers to be condensed in the atmosphere of SDSS1411B. 

\begin{figure*}[t]
    \centering
    \includegraphics[width = 1\textwidth]{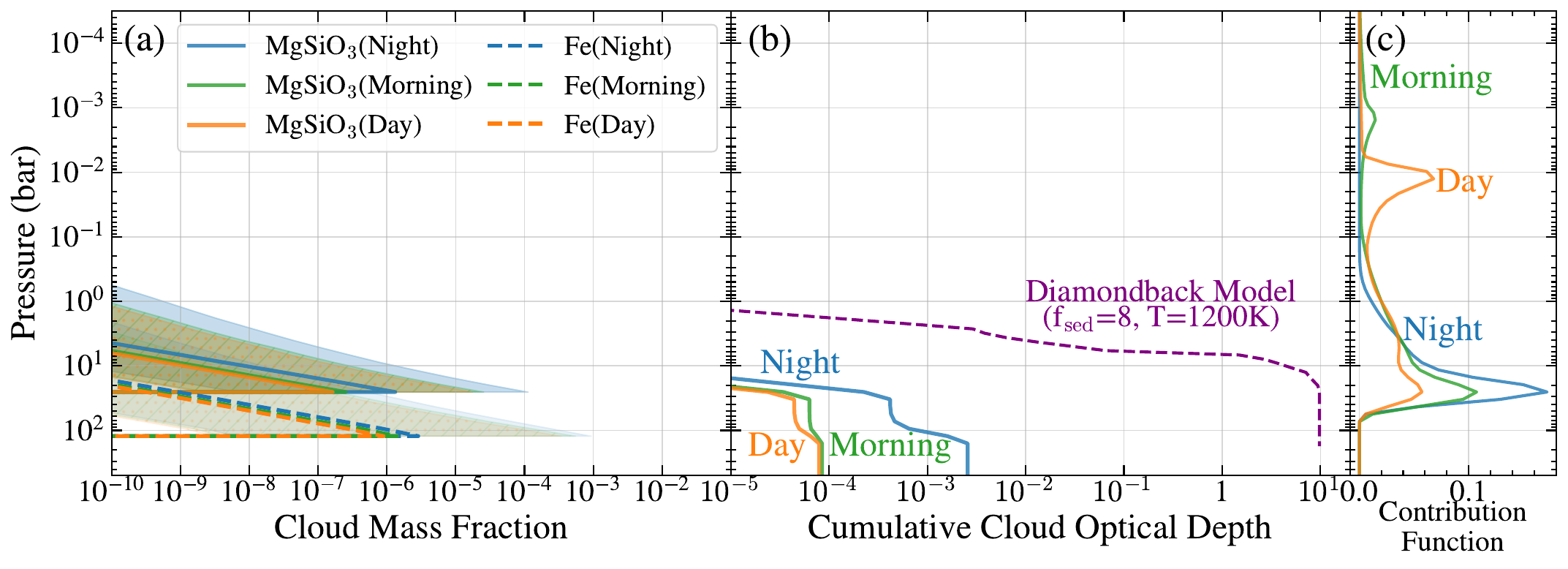}
    \caption{Panel (a): Cloud abundance profiles of MgSiO$_3$ (solid) and Fe (dashed) for the Day (orange), Morning (green), and Night (blue) phases from the cloudy retrieval model with free CO$_2$ chemistry, with shaded regions indicating the abundance uncertainties. MgSiO$_3$ condensates form near $\sim$30~bar, whereas Fe clouds form at deeper atmospheric pressures of $\sim$100~bar. Panel (b): Cumulative cloud optical depth at 3~$\upmu$m for each longitude. The purple dashed line shows the cloud optical-depth profile of the Sonora Diamondback model with $T_{\rm eff}=1200$~K and $f_{\rm sed}=8$ for comparison. Panel (c): Corresponding contribution functions integrated over 0.6--5.3~$\upmu$m.}
    \label{fig:cloud_abun}
\end{figure*}

Rather than the inversion being driven by such optical absorbers, we hypothesize that SDSS1411B's temperature is driven by UV absorbers. More than 60$\%$ of the irradiation from its $T_{\rm eff}\sim12{,}000$~K WD host is emitted at wavelengths shorter than the optical, with the peak flux at about 0.25~$\upmu$m. Flux at such wavelengths is preferentially absorbed at lower pressures in the object's atmosphere \citep{Lothringer18} and is more capable of driving upper atmosphere heating. This is supported by theoretical modeling of planets around different host star types \citep{Lothringer2019}, as well as observations of such planets \citep[e.g,][]{Yan2020}. At some point, the atmosphere at low pressures can become hot enough that H$_2$O is dissociated. The dissociation can further drive a temperature inversion due to reduced ability of the atmosphere to cool via radiation \citep{Parmentier18, Lothringer18}.

This scenario is supported by 1D PHOENIX atmosphere models including irradiation. Using the same setup as the models from \citep{Lothringer2020}, we irradiate SDSS1411B with a $T_{\rm eff}\sim13000$~K WD host star and assume dayside heat redistribution. We also include rainout, which prevents TiO, VO, Fe, and other metal species from existing in the gas phase at low pressures. Figure~\ref{fig:1d_model} shows that this model exhibits many of the same qualitative features as the observed dayside spectrum of SDSS1411B, including a strong temperature inversion, as well as the mix of absorption and emission features. We also compute an identical model but changed the irradiation spectrum to represent a $T_{\rm eff}\sim6000$~K main-sequence host star. In this comparison model, no strong inversion forms and all major molecular features are found to be in absorption, supporting our hypothesis of a previously unobserved UV-driven temperature inversion in SDSS1411B.

\begin{figure*}[]
    \centering
    \includegraphics[width = 1\textwidth]{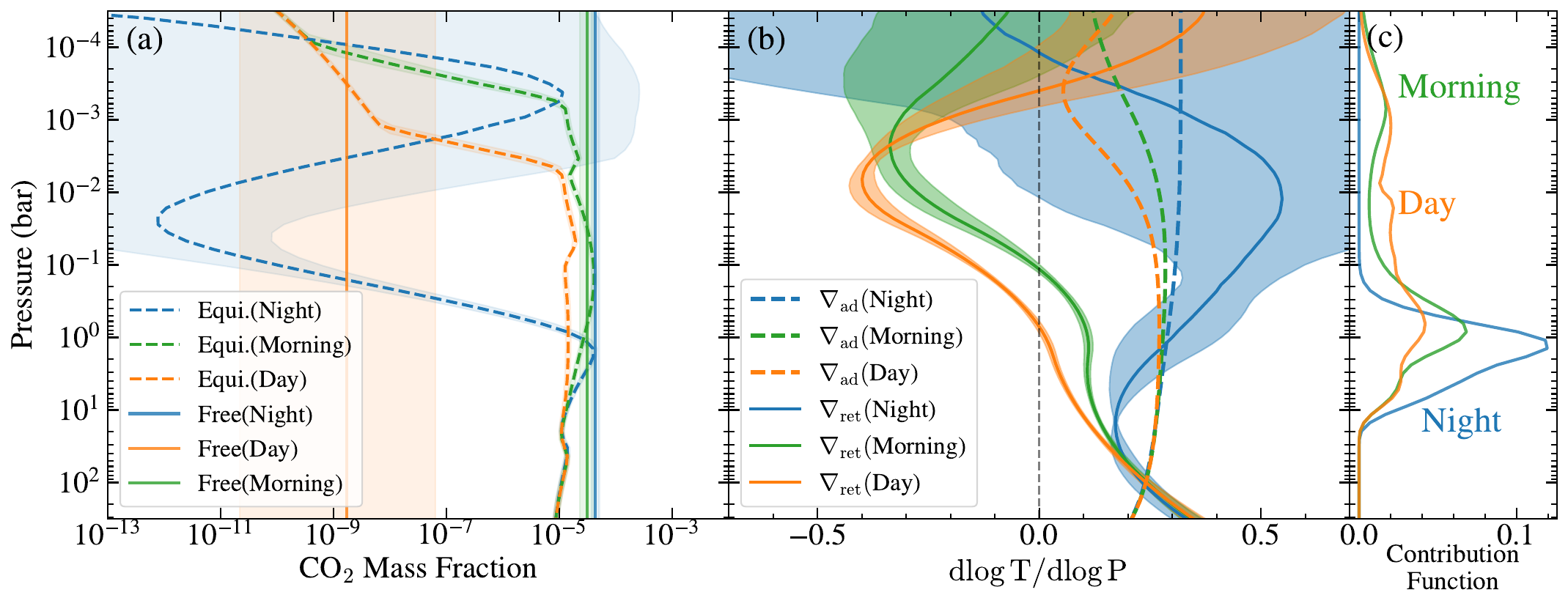}
    \caption{Panel (a): CO$_2$ abundance profiles for retrievals with equilibrium chemistry (dashed lines) and  free CO$_2$ chemistry (solid lines). The best-fit free-CO$_2$ retrieval favors a lower CO$_2$ abundance on the dayside than on the morning- and nightsides, while the nightside CO$_2$ abundance exceeds that predicted by the equilibrium model. Panel (b): Temperature-gradient profiles derived from the retrieved T--P profiles at each longitude (Figure~\ref{fig:tp_tau}a). The dashed line indicates the adiabatic gradient adopted from \cite{Parmentier15}. The nightside profile exceeds the adiabatic gradient over $\sim$0.001--1~bar, suggesting potential convective instability. Panel (c): Corresponding contribution functions integrated over 4-5.3 $\upmu$m, including the CO$_2$ band. }
    \label{fig:co2_abun}
\end{figure*}

\subsection{Condensate Clouds}\label{sec:dis_cloud}
The atmospheric retrieval analysis in Section~\ref{sec:retrieval} well reproduces the observed SDSS1411B spectra without cloud components. Including cloud opacity in the retrieval modeling does not provide a significant improvement in the spectral fit shown in Figure~\ref{fig:retrieval_spec}(a), providing lower marginal evidence. This perhaps unexpected absence of clouds can be understood considering additional modeling results: The retrieved T--P profile (Figure~\ref{fig:tp_tau}b) suggests that condensates may form in the deep atmosphere (between $\sim$30 and 300~bar pressures), whereas most of our observations probe pressures lower than 30~bar. Furthermore, the condensate abundance profile of the cloudy model (Figure~\ref{fig:cloud_abun}a) indicates a negligible cloud layer near $\sim$30~bar, with a cloud-base condensate mass fraction less than $\sim10^{-5}$. The low cumulative cloud optical depth, $\lesssim0.01$ at 100~bar, further supports the negligible effect of clouds on the observed spectra of SDSS1411B (Figure~\ref{fig:cloud_abun}b). Together, these retrieval results suggest that cloud opacity is not required to reproduce the observed spectra of SDSS1411B.

Given the low nightside temperature, this may be surprising. We speculate that the strong external irradiation may impact cloud formation in SDSS1411B. In this scenario, irradiation would heat the upper atmosphere and evaporate condensates. This is shown in Figure~\ref{fig:tp_tau}(b), where the heated dayside T--P profile does not cross the MgSiO$_3$ condensation curve beyond the 1$\upsigma$ uncertainty, suppressing cloud formation on the dayside. This model result is consistent with the GCM studies of hot Jupiters predicting that refractory condensates from the cooler nightside can evaporate as they are transported toward the hotter dayside (e.g., \citealt{Parmentier13,Parmentier16, Mukherjee26}). In addition, the nightside cloud distribution may be affected by day-to-night material transport and disequilibrium chemistry (e.g., \citealt{Mendonca18,Bell24}). Such irradiation-driven processes could alter the abundances and chemical phases of cloud-forming species, potentially further suppressing the formation or survival of extensive and homogeneous cloud layers.


One caveat in the above scenario is that the observed spectrum does not directly rule out the presence of clouds on the nightside. Rather, our results show that the observed spectra can be adequately reproduced by a cloudless model and that including cloud opacity is not statistically favored, because the additional cloud model parameters do not provide a significant improvement in the spectral fit. Future mid-infrared observations, such as with JWST/MIRI, will be important for testing the presence of clouds and constraining their properties in SDSS1411B by probing the characteristic cloud features at $\gtrsim$ 5~$\upmu$m (e.g., \citealt{Grant23,Heinke26}).

\subsection{CO$_2$ Disequilibrium Chemistry}\label{sec:dis_cloud}

Our retrieval analysis statistically favors the model with a free, vertically constant CO$_2$ mass fraction (see Section~\ref{sec:retrieval}). To further investigate potential CO$_2$ disequilibrium chemistry, we compared the CO$_2$ abundance profiles of the best-fit cloudless model with free CO$_2$ chemistry and the fully equilibrium cloudless model in Figure~\ref{fig:co2_abun}(a). The free-CO$_2$ model retrieves a lower CO$_2$ abundance on the dayside than on the morning and nightside hemispheres. This result remains broadly consistent with the equilibrium model because the dayside contribution function across the CO$_2$ band is concentrated in the upper atmosphere (Figures~\ref{fig:2d_contri} and \ref{fig:co2_abun}c), where the equilibrium CO$_2$ abundance is already low. In contrast, the free-CO$_2$ model retrieves a higher CO$_2$ abundance on the nightside than predicted by the equilibrium model, which alleviates the discrepancy near the 4~$\upmu$m CO$_2$ feature (Figure \ref{fig:retrieval_spec}a).

In fact, there are reasons why CO$_2$ may be out of equilibrium: For example, the irradiation by the WD may reduce the CO$_2$ abundance of the BD through thermal and photochemical dissociation. Similar to the thermal dissociation of molecules in highly irradiated hot Jupiters (e.g., \citealt{Parmentier18,Gandhi20}), the high temperatures and prominent temperature inversion on the dayside can drive CO$_2$ dissociation and deplete its abundance in the upper atmosphere. In addition, strong external UV irradiation from the WD can photochemically dissociate CO$_2$ into CO in the dayside upper atmosphere \citep{Tian09,Zahnle09}. Together, these irradiation-driven processes can drive the atmospheric composition away from thermochemical equilibrium and may contribute to the low CO$_2$ abundance retrieved on the dayside.

Similarly, vertical convection and horizontal advection may also contribute to the departure of the nightside CO$_2$ abundance from chemical equilibrium. To assess potential convective instability in SDSS1411B, we calculated the temperature gradients of the retrieved T--P profiles (Figure~\ref{fig:tp_tau}a) and compared them with the adiabatic gradient from \cite{Parmentier15}, based on a fit to the high-pressure equation of state of \cite{Saumon95}, $\nabla_{\rm ad}\sim0.32-0.1(T/3000~{\rm K})$. The retrieved temperature gradient, $\nabla_{\rm ret}$, is smaller than $\nabla_{\rm ad}$ at $\sim$1--100~bar, consistent with the HST-based modeling results of \cite{Lew+22}. In contrast, the nightside profile exhibits $\nabla_{\rm ret}>\nabla_{\rm ad}$ over $\sim10^{-3}$--1~bar, indicating potential convective instability in the same pressure range where the retrieved CO$_2$ abundance departs from equilibrium. Although the nightside temperature gradients remain highly uncertain in the upper atmosphere, this stability analysis suggests that atmospheric dynamics may contribute to the observed disequilibrium chemistry in SDSS1411B.

\subsection{Thermal and Radius Evolution of SDSS1411B}

Our inferred cooling age for SDSS1411B (Section~\ref{sec:e_transport}) is marginally consistent with the $\gtrsim3$~Gyr age range estimated in previous studies \citep{Littlefair+14,Lew+22}, although discrepancies remain among age estimates derived using different methods. The $\gtrsim3$~Gyr constraint from \citet{Littlefair+14} is based on the BD mass--radius relation, yet their inferred BD radius is more than $3\sigma$ smaller than predicted by the 3~Gyr \citet{Baraffe03} isochrone. Similarly, the Sonora evolutionary models (Section~\ref{sec:e_transport}; \citealt{Marley21}) predict a larger radius than the value derived in Section~\ref{sec:sys_param}. This compact radius is also inconsistent with the radius inflation predicted for irradiated BDs \citep{Mukherjee26b}. The discrepancy may reflect differences in atmospheric properties or irradiation conditions \citep{Beatty26}, including the much bluer incident spectrum of the WD compared with the cooler main-sequence stellar irradiation considered by \citet{Mukherjee26b}. Furthermore, the rapid rotation of SDSS1411B may suppress vertical energy transport, potentially limiting irradiation-driven radius inflation \citep{Sainsbury-Martinez21, Sainsbury-Martinez24}. On the other hand, the inferred radius of SDSS1411B, $\sim0.72~R_{\rm J}$, lies within the observed radius distribution of weakly irradiated transiting BDs \citep{Beatty26}. The SDSS1411 system may have completed common-envelope evolution on a relatively short timescale, potentially $\lesssim$ 10$^3$ yr \citep{Izzard12, Jenkins26}, after which WD irradiation may have only a limited effect on thermal and radius evolution of the BD during its post-common-envelope phase.


Despite the inferred BD cooling age from well-constrained L$\rm{_{int}}$, constraining the total system age remains challenging. Common-envelope evolution and subsequent tidal interactions may alter the BD radius and thermal state \citep{Rasio96,Weinberg17,Lew+22}. In addition, accretion simulations coupled with stellar evolution models predict that a substellar companion may accrete substantial mass during common-envelope evolution \citep{Chamandy18,Jenkins26}, potentially obscuring its pre-common-envelope thermal history. Studies of close WD–BD binaries therefore suggest that the BD cooling age may provide only a lower limit on the system age if the BD was significantly reheated during common-envelope evolution. \citep{Burleigh06,Zorotovic22}. Future detailed modeling of common-envelope and tidal evolution, combined with system parameters constrained by phase-resolved bolometric observations, will be essential for fully understanding the energy budget and evolutionary history of the SDSS1411 system. 


\subsection{SDSS1411B and Other Irradiated Brown Dwarfs}

We compare SDSS1411B with other irradiated BDs to examine three key properties: (1) strong temperature inversion in the dayside, (2) the cloudless hemispheres, and (3) CO$_2$ disequilibrium chemistry. 

The strong dayside temperature inversion identified in Sections~\ref{sec:results} and \ref{sec:atm_model} places SDSS1411B as the least strongly irradiated BD known to exhibit a significant dayside inversion. Other irradiated BDs with reported dayside inversions, including EPIC2122B, SDSS1557B, and WD0137B \citep{Lothringer24,Amaro24}, receive incident irradiation fluxes of $\gtrsim10^{9}~{\rm erg~s^{-1}~cm^{-2}}$, higher than that of SDSS1411B, $\sim$ $10^{8.6}~{\rm erg~s^{-1}~cm^{-2}}$. In contrast, less strongly irradiated systems such as NLTT5306B, GD1400B, and ZTF0038B \citep{Amaro23,Amaro25,BroskiLaing2026} show no evidence of dayside temperature inversions in their HST or JWST spectra. SDSS1411B therefore occupies an important irradiation regime for investigating the conditions under which stellar irradiation drives thermal inversions, particularly the role of short-wavelength heating in the upper atmosphere.   

Previous studies revealed a substantial diversity in cloud distribution among WD-irradiated BDs. Atmospheric modeling of HST observations of SDSS1557B, which receives a higher irradiation flux than SDSS1411B, favors a cloudy nightside and cloud-free dayside \citep{Amaro24}. On the other hand, HST spectra of the highly irradiated EPIC2122B are well reproduced without including cloud opacity \citep{Lothringer24}. Among more weakly irradiated systems, WD1032+011B exhibits a cloudy nightside and clear dayside \citep{French+24}, suggesting dayside cloud dissipation by stellar irradiation, whereas GD1400B, with the weakest irradiation flux, shows evidence for global cloud coverage on both hemispheres \citep{Farihi04, Farihi05,Amaro25}. These contrasting results indicate no clear correlation between cloud coverage and irradiation strength, suggesting that additional parameters, such as BD mass, effective temperature, or spectral type, may also regulate cloud formation. Systematic cloud analyses based on broad-wavelength spectroscopy of a larger sample of irradiated BDs will therefore be essential for determining how external irradiation influences phase-dependent cloud formation.

\citet{BroskiLaing2026} presented the first JWST/NIRSpec phase curve of the irradiated BD ZTF0038B, which orbits an $\sim$11,000 K WD at a separation of $\sim$2 R$_{\odot}$ and therefore receives a lower irradiation flux ($\sim10^{7.6}~{\rm erg~s^{-1}~cm^{-2}}$) than SDSS1411B ($\sim10^{8.6}~{\rm erg~s^{-1}~cm^{-2}}$). Their observations revealed asymmetric CO$_2$ absorption-band phase curves and nearly homogeneous CO$_2$ abundances across orbital phase. Similar to SDSS1411B, ZTF0038B shows departures from chemical-equilibrium predictions, potentially driven by vertical convection and horizontal advection \citep{Bell24,Parmentier26}. However, unlike SDSS1411B, ZTF0038B exhibits relatively homogeneous CO$_2$ abundances across longitude, possibly reflecting more efficient horizontal advection or weaker external irradiation that is insufficient to drive substantial thermal or photochemical dissociation on the dayside.


Studying a broader sample of irradiated BDs will provide a foundation for understanding irradiation-driven atmospheric dynamics and disequilibrium chemistry in the high-gravity, rapidly rotating, and high-internal-luminosity regime. Analysis of the three remaining irradiated BDs in JWST program GO-4967, WD1202, WD0137, and WD0137 (French et al., Broski-Laing et al., and Shin et al., in prep.), will enable a more systematic investigation of the atmospheric physics across different irradiation environments. Future discoveries of additional WD--BD systems through optical photometric surveys such as ZTF and ATLAS (e.g., \citealt{vanRoestel21,Smith20}) and infrared surveys (e.g., \citealt{Wilson23,Gagne26}) will further expand the sample and improve its statistical homogeneity.

\section{Conclusions}\label{sec:conclusion}
We presented time-resolved JWST/NIRSpec PRISM spectrophotometry of SDSS1411B, which occupies an intermediate irradiation regime among five close WD--BD binaries in the GO-4967 program (PI: Yifan Zhou). Our atmospheric analysis reveals atmospheric dynamics, thermal inversion, and disequilibrium chemistry in this strongly UV-irradiated BD. The key findings of this study are as follows:

\begin{itemize}
    \item The SDSS1411 system shows an average variability amplitude of $\sim$14$\%$ of the total flux, increasing from a trough-to-peak variation of $\sim$2$\%$ at 0.6--1.47~$\upmu$m to $\sim$54$\%$ at 4.40--5.20~$\upmu$m.

    \item The second-order Fourier model successfully reproduces the observed BD light curve. Wavelength bins with significant correlations between Fourier amplitudes of stationary and time-varying components coincide with wavelength regions dominated by H$_2$O, CH$_4$, and CO absorption features at $\sim$1-2.4 and $\sim$4 $\upmu$m, suggesting larger flux variations in the upper atmosphere than in deeper atmospheric layers.

    \item The BD brightness temperatures range from $\sim$600 to 2,000~K on the nightside and from $\sim$1,500 to 2,000~K on the dayside. The spectral regions near $\sim$3 and 5 $\upmu$m exhibit the lowest nightside brightness temperatures ($\sim$600 K) and probe relatively low-pressure atmospheric layers ($\sim$1 bar) with short radiative timescales ($\lesssim$2 hr), consistent with the large flux and brightness temperature variations observed at these wavelengths.

    
    \item The BD nightside spectrum is inconsistent with the non-irradiated 1D forward models, showing a deviation from the equilibrium assumptions possibly due to the irradiation effects. In contrast, the atmospheric retrieval analysis successfully reproduces the nightside BD spectra with a cloud-free model and free CO$_2$ chemistry. The retrieved nightside model shows a monotonic T--P profile, showing departure from that of the thermochemical-equilibrium model at $\lesssim$ 1 bar.     
    
    \item The BD dayside spectrum is not well fit by non-irradiated 1D forward models. In contrast, the atmospheric retrieval analysis provides a cloudless best-fit model spectrum, in agreement with the observed dayside spectra. The retrieved dayside T--P profile shows a prominent temperature inversion near $\sim$0.1~bar, suggesting that SDSS1411B is the lowest-temperature irradiated object known to exhibit a thermal inversion.

    \item The phase-resolved spectra of SDSS1411B can be adequately reproduced by both cloudy and cloudless models, although the retrieval statistically favors the cloudless model. The best-fit cloudy model expects negligible condensate mass fractions ($\lesssim$ 10$^{-4}$) and low cumulative cloud optical depths ($\lesssim$ 10$^{-2}$) across all hemispheres, indicating that clouds have little effect on the observed spectra.

    \item The free-CO$_2$ chemistry model provides a better fit to the observed spectra than the equilibrium-chemistry model, suggesting irradiation-driven disequilibrium chemistry in SDSS1411B, potentially including thermal and photochemical CO$_2$ dissociation on the dayside and vertical mixing on the nightside.

    \item The bolometric correction of the observed phase-resolved BD spectra provides effective temperatures of $1703\pm$58 K for the dayside and $1272\pm$48 K for the nightside. Our energy transport analysis indicates inefficient heat redistribution between the two hemispheres, with a transport efficiency of $<0.1$. The inferred internal luminosity constrains the BD cooling age to $\sim$3~Gyr, while the derived BD radius is significantly smaller than predicted by evolutionary models.

    \item Comparisons between SDSS1411B and other irradiated BDs in terms of (1) dayside temperature inversions, (2) cloud coverage, and (3) CO$_2$ disequilibrium chemistry establish SDSS1411B as an important laboratory in a unique irradiation regime for investigating how external irradiation shapes substellar atmospheres.

\end{itemize}

We present high-quality, phase-resolved JWST spectrophotometry of the close WD--BD binary SDSS1411. Our multi-phase joint retrieval framework successfully reproduces the observed spectra and provides robust constraints on the atmospheric properties across different orbital phases. We identify a prominent dayside temperature inversion, largely cloud-free hemispheres, and CO$_2$ disequilibrium chemistry. These results highlight the potential of JWST observations of close WD--BD systems, providing new insights into how strong external irradiation shapes substellar atmospheres.

\begin{acknowledgments}
We thank the anonymous referee for constructive comments that improved the manuscript. This work is based on observations made with the NASA/ESA/CSA James Webb Space Telescope. National Aeronautics and Space Administration supported this work through a grant under the JWST-GO-4967 program from the Space Telescope Science Institute and through agreement No. 80NSSC21K0593 for the program ``Alien Earths.'' The data were obtained from the Mikulski Archive for Space Telescopes at the Space Telescope Science Institute, which is operated by the Association of Universities for Research in Astronomy, Inc., under NASA contract NAS 5-03127 for JWST. The specific observations analyzed in this study can be accessed via \dataset[doi: 10.17909/zrd9-d176]{https://doi.org/10.17909/zrd9-d176}. SX is supported by the international Gemini Observatory, a program of NSF NOIRLab, which is managed by the Association of Universities for Research in Astronomy (AURA) under a cooperative agreement with the U.S. National Science Foundation, on behalf of the Gemini partnership of Argentina, Brazil, Canada, Chile, the Republic of Korea, and the United States of America. AI tools were used to suggest language improvements within the manuscript.

\end{acknowledgments}

\begin{contribution}
LS reduced and analyzed the data, carried out the forward modeling and retrieval studies, interpreted the model results, and drafted the manuscript. DA, YZ, and DB-L contributed significantly to the data analysis and modeling strategy, interpretation, and provided comments to the manuscript draft. YZ led the proposal and planned the observations. DA advised LS. JL contributed significantly to the atmospheric modeling studies and drafted the discussion of the dayside model fits. SC, SX, and BL contributed to the characterization and interpretation of the WD-BD system parameters and provided comments to the manuscript draft. MSM, LM, XT, JF, and VP contributed to the atmospheric modeling and interpretation of the BD spectrum. XT, JF, and VP also reviewed the manuscript and provided comments.
\end{contribution}

\bibliography{sample701}{}
\bibliographystyle{aasjournalv7}

\end{document}